\documentclass[12pt,a4paper]{book}
\usepackage{Wiley-AuthoringTemplate}
\usepackage[numbers]{natbib}
\usepackage[caption=false,font=normalsize,labelfont=sf,textfont=sf]{subfig}
\usepackage[]{algorithmicx}
\usepackage{algpseudocode,algorithm}
\usepackage{multirow}

\usepackage{glossaries}

\RequirePackage{times}

\newcommand{\bm}[1]{\mbox{\boldmath{$#1$}}}

\DeclareMathOperator*{\argmin}{argmin} 
\makeindex

\begin{document}
\singlespacing

\setcounter{chapter}{5}
\chapter[Deep Learning for Sparse Arrays]{Sparse Array Design for Direction Finding using Deep Learning\protect\footnote{K. V. M. acknowledges support from the National Academies of Sciences, Engineering, and Medicine via Army Research Laboratory Harry Diamond Distinguished Fellowship.} \vspace{-24pt} }

\author*[1]{Kumar Vijay Mishra}
\author[2]{Ahmet M. Elbir}
\author[3]{Koichi Ichige}

\address[1]{\orgdiv{Computational and Information Sciences Directorate (CISD)}, 
\orgname{United States DEVCOM Army Research Laboratory}, 
\postcode{20783}, \countrypart{Adelphi}, 
     \city{Maryland}, \country{USA}}%

\address[2]{\orgdiv{Interdisciplinary Centre for Security, Reliability and Trust (SnT)}, 
\orgname{University of Luxembourg}, 
\postcode{1855}, \street{Av. John F. Kennedy}, \country{Luxembourg}}%

\address[3]{\orgdiv{Department of Electrical and Computer Engineering}, 
\orgname{Yokohama National University}, 
\postcode{240-8501},  
     \city{Yokohama}, \country{Japan}}%
     
\address*{Corresponding Author: Kumar Vijay Mishra; \email{kvm@ieee.org}}

\maketitle

\vspace{-24pt}
\begin{abstract}{Abstract}
In the past few years, deep learning (DL) techniques have been introduced for designing sparse arrays. These methods offer the advantages of feature engineering and low prediction-stage complexity, which is helpful in tackling the combinatorial search inherent to finding a sparse array. In this chapter, we provide a synopsis of several direction finding applications of DL-based sparse arrays. We begin by examining supervised and transfer learning techniques that have applications in selecting sparse arrays for a cognitive radar application. Here, we also discuss the use of meta-heuristic learning algorithms such as simulated annealing for the case of designing two-dimensional sparse arrays. Next, we consider DL-based antenna selection for wireless communications, wherein sparse array problem may also be combined with channel estimation, beamforming, or localization. Finally, we provide an example of deep sparse array technique for integrated sensing and communications (ISAC) application, wherein a trade-off of radar and communications performance makes ISAC sparse array problem very challenging. For each setting, we illustrate the performance of model-based optimization and DL techniques through several numerical experiments. We discuss additional considerations required to ensure robustness of DL-based algorithms against various imperfections in array data.
\end{abstract}


\section{Introduction}
\label{sec:Introduciton}
Phased sensor arrays have become widely prevalent in various applications such as radar, sonar, communications, acoustics, sounding, and ultrasound \cite{elbir2022Nov_Beamforming_twentyfive,shenoy1994phased,frank2008advanced,herd2015evolution}. They allow electronic beam steering without requiring any mechanical motion, providing high spatial selectivity and the ability to reject interference. However, according to the Nyquist-Shannon theorem, the array must allow for at least two signal samples in a single spatial period, which is the same as the array's operating wavelength \cite{haupt2015timed}. Otherwise, \textit{spatial aliasing} occurs resulting in multiple main-lobes appearing in the beampattern, which reduces the directivity and affects the accuracy of estimating bearings or directions-of-arrival (DoAs) of unknown sources or targets \cite{johnson1982application}. To avoid this, conventional phased sensor arrays employ a uniform linear array (ULA), wherein sensors must be separated \textit{at most} by half-wavelength spacing. However, as the number of sensors increases, the associated complexity, size, and cost of arrays become unacceptable. As a result, there is growing interest in thinned or sparse sensor arrays, which offer significantly reduced hardware \cite{linebarger1993difference,haupt1994thinned,mishra2017high}.

Uniformly removing sensors from a ULA produces \textit{grating lobes} or copies of the main lobe. This effect can be suppressed by removal of sensors at random locations giving rise to \textit{thinned} or \textit{sparse} arrays \cite{lo1964mathematical,agrawal1972mutual}. However, randomly thinned arrays are accompanied by increased sidelobe levels, which hampers direction finding. Further, the number of elements in an array determines its degrees-of-freedom (DoFs) or the number of sources whose bearings can be determined by the sensor array. A full or filled sensor array with N elements has more DoFs than a sparse array with only 
\(\mathcal{O}(\sqrt{N})\) sensors \cite{haupt1994thinned,elbir2022Nov_Beamforming_twentyfive}. But, with suitable parameter recovery algorithms \cite{dspCoprime1,mishra2017high,sedighi2019optimum,lv2023co}, sparse arrays have been shown to exhibit negligible performance degradation and reduced mutual coupling in direction finding \cite{boudaher2017mutual,superNestedArray,liu2017hourglass} and spatial filtering \cite{elbirQuantizedCNN2019}. Most of these algorithms employ sparse recovery techniques, including compressed sensing (CS) \cite{foucart2013mathematical}.

\subsection{Prior Art and Historical Notes}
Sparse array literature could be traced back to the seminal works of Y. T. Lo \cite{lo1964mathematical,agrawal1972mutual}, who studied the probabilistic properties (e.g. the probability of the array pattern crossing a specified sidelobe level) of random arrays. Other studies published around the same time \cite{unz1962nonuniform,ishimaru1965thinning,Sandler1960Equivalences} focused on approximating the analytic expressions of desired antenna patterns with few nonuniformly-spaced elements but these were optimal only for smaller arrays. This was followed by sparse array designs using vastly different procedures such as dynamic programming \cite{skolnik1964dynamic}, iterative least squares \cite{redlich1973iterative}, nonlinear optimization \cite{schjaer1976synthesis}, genetic algorithms \cite{haupt1994thinned}, and simulated annealing (SA) \cite{murino1996synthesis}. These techniques are usually limited by slow speed of optimization, high sidelobes, non-optimality for large arrays, and high dependence of final element placement based on the choices made in previous iterations. 

Other geometries of non-uniformly-spaced sparse arrays that have low sidelobes were also proposed; see, e.g., \cite{haupt2010antenna} for an overview, and references therein. For example, \cite{kim1986fractal} describes a fractal-based quasi-random array that combines properties of both random and periodic arrays such that its peak sidelobe level (PSL) is lower than a purely random array. A space-tapered array \cite{lo1966study,willey1962taper} places antennas with increasing inter-element spacing as one moves away from the aperture's center thereby achieving low PSL but higher integrated sidelobe levels (ISL) than random arrays. 

A sparse array is hole-free or \textit{restricted} if its difference coarray is a ULA, otherwise it is a \textit{general} array. A sparse array with $N$ sensors and the aperture defined by the integer grid $[-M,M]$ may also be characterized by its \textit{redundancy ratio} $R=N(N-1)/(2M) \geq 1$, which is the number of pairs of antennas divided by the aperture \cite{moffet1968minimum}. Using this metric, zero redundancy arrays (ZRAs) \cite{arsac1955nouveau}, minimum redundancy arrays (MRAs) \cite{moffet1968minimum}, and low redundancy arrays (LRAs) \cite{camps2001synthesis} were proposed. These minimize spatial aliasing while maintaining a reasonably constant ISL. The ZRAs have $R=1$ but they exist for only $N \leq 4$ \cite{bracewell1966optimum}. the MRAs, which achieve the lowest $R$, select the missing elements such that the resulting sparse array has all possible inter-element spacings of the full array. Leech's bounds \cite{leech1956representation} for MRAs stipulate $1.217 \leq R \leq 1.674$ for $N \rightarrow \infty$. Optimal solutions for $N \leq 11$ were provided in \cite{leech1956representation} and, using exhaustive computer-aided search, expanded later to $N \leq 26$ \cite{schwartau2021large}. For larger $N$, the procedure becomes increasingly complex \cite{ishiguro1980minimum} and there have been efforts to meet the Leech's lower bound through LRAs \cite{camps2001synthesis}. These problems have led to the popularity of more structured sparse designs such as co-prime arrays \cite{coprime_PPV_Vaidyanathan2010Oct,SQin2015,WZheng2021}. While these configurations provide a closed form of sensor positions and offer enhanced DoFs for parameter estimation, they are not applicable to arbitrary number of antenna elements.

\subsection{Learning-Based Approaches} 
In general, searching for an optimal sparse sensor array is a combinatorial problem \cite{moffet1968minimum}, whose computational complexity increases with the number of sensors. Since a closed-form solution is difficult to come by, several sub-optimal (although mathematically tractable) solutions have been proposed \cite{kozick1991linear,superNestedArray,sedighi2019optimum,antennaSelectionForMIMO,antennaSelectionViaCO, antennaSelectionKnapsack}. Recently, there has been a growing interest in utilizing learning-based approaches in sparse sensor communications \cite{elbirQuantizedCNN2019,elbir2019robust} and signal processing \cite{elbir2019cognitive,deepLearning4SignalProcessing}. 

Deep learning (DL) has been shown to offer better computational efficiency than a combinatorial search \cite{elbirQuantizedCNN2019,elbir2019deepursi}.
DL has been extremely effective in addressing challenging problems such as speech recognition, visual object recognition, and language processing \cite{deepLearningScience, deppLearningRepresetation}. It offers advantages such as low computational complexity while tackling optimization-based or combinatorial search problems, and the ability to generate new features from a limited set of features available in a training set \cite{svmDoAEst1, deepLearningScience}. 

	\begin{figure}[t]
		\centering
		\includegraphics[width=0.80\textwidth]{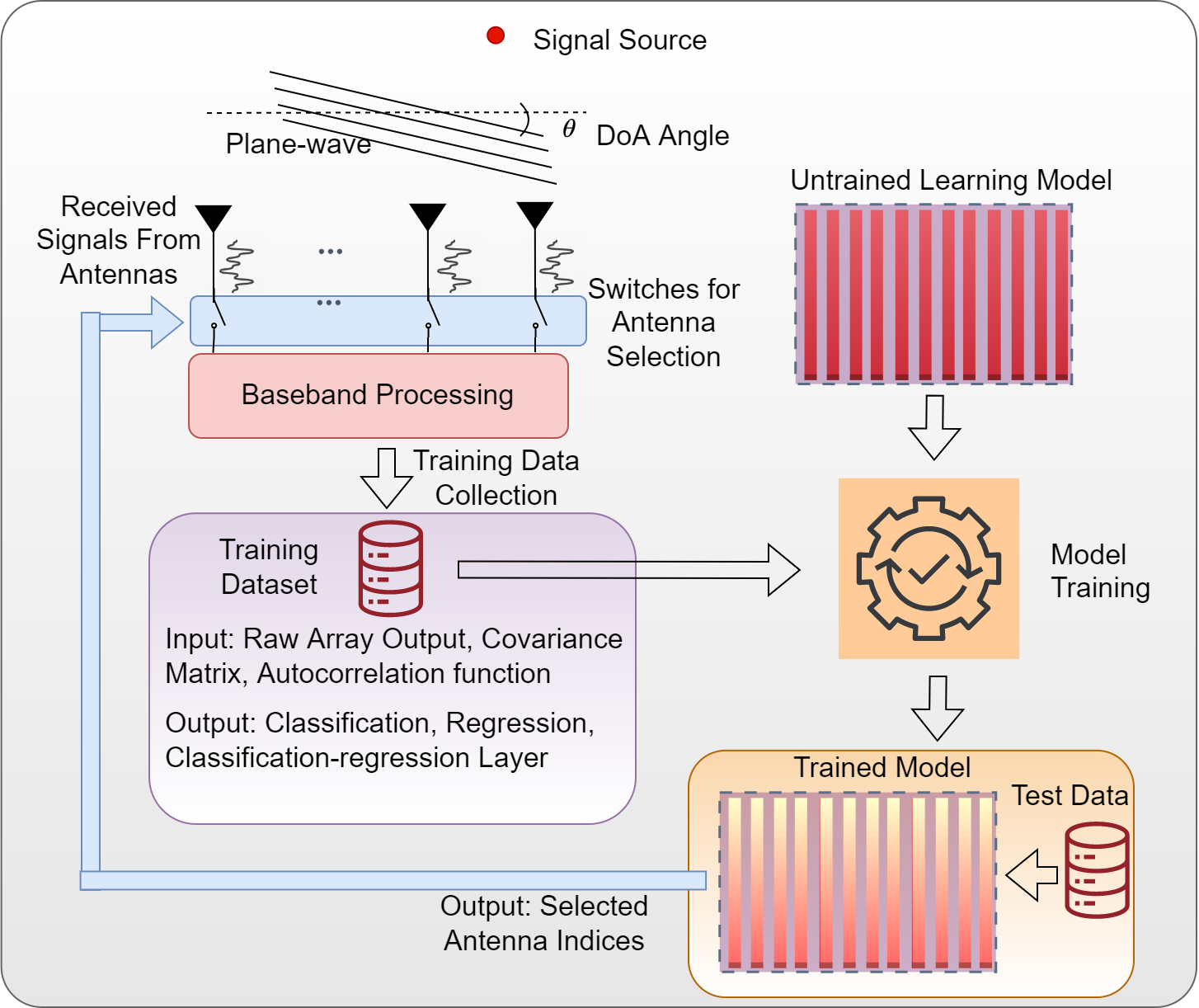} 
		\caption{Illustration of DL-based antenna selection. }
		\label{figAntennaSelectionDL}
	\end{figure}
	
 In a typical DL-based sparse array design (Fig.~\ref{figAntennaSelectionDL}), the antenna array data is fed into the DL network, which yields the indices of the selected antennas at the output. To train the DL model, the input datasets vary and may comprise of the raw array output~\cite{elbirQuantizedCNN2019}, covariance matrix~\cite{elbir_DL_MUSIC} or the autocorrelation function~\cite{df_featureEngineeringKulkarni2021Oct} of the array data. The output layer of the model is usually a classification layer, wherein each class corresponds to a possible set of subarray indices. In some case, the output is a regression layer, whose values are either $0$ or $1$ indicating the (un)selected antennas~\cite{elbir2019cognitive}.

The rest of the chapter is organized as follows: In the next section, we review general DL design procedures for sparse array design and direction-of-arrival (DoA) estimation. Then, Section~\ref{sec:ant_sel}) describes a major use case of this technique in cognitive radar, where a dynamic transmit antenna selection is required for the changing target environment. A DL-based approach is able to yield sparse (sub-)arrays in real-time for this application. When insufficient training data are available for a new antenna geometry, then Section~\ref{sec:transfer} presents a transfer learning (TL) approach to utilize a trained network from a different antenna configuration. For planer or two-dimensional (2-D) large sparse arrays, usually a combination of SA and DL may be more appropriate, as detailed in Section~\ref{sec:2d}. We then consider DL for sparse array applications in wireless communications by combining antenna selection with massive multiple-input multiple-output (MIMO) beamforming (Section~\ref{sec:beamforming}) and integrated sensing and communications (ISAC) \cite{mishra2019toward} (Section~\ref{sec:isac}). Finally, we summarize our key observations and unaddressed challenges in Section~\ref{sec:summ}.

\textit{Notations:} Throughout the chapter, we reserve boldface lowercase and uppercase letters for vectors and matrices, respectively. The $i$th element of a vector $\textbf{y}$ is $y(i)$ while the $(i,j)$th entry of the matrix $\textbf{Y}$ is $[\textbf{Y}]_{i,j}$. We denote the transpose and Hermitian by $(\cdot)^T$ and $(\cdot)^H$, respectively. The functions $\angle\left\lbrace \cdot \right\rbrace$, $\operatorname{\mathbb{R}e}\left\lbrace \cdot \right\rbrace$ and $\operatorname{\mathbb{I}m}\left\lbrace \cdot \right\rbrace$ designate the phase, real and imaginary parts of a complex argument, respectively. The combination of selecting $K$ terms out of $M$ is denoted by $\left( \begin{array}{c} M \\ K \end{array}\right) = \frac{M !}{K!(M-K)!}$. The Hadamard (point-wise) product is written as $\odot$. The functions $\text{E}\left\lbrace \cdot \right\rbrace $ and $\text{max}$ give the statistical expectation and maximum value of the argument, respectively. The notation $x \sim \textrm{u}\{[u_l,u_u]\}$ means a random variable drawn from the uniform distribution over $[u_l,u_u]$ and $x \sim \mathcal{CN}(\mu_x,\sigma_x^2)$ represents the complex normal distribution with mean $\mu_x$ and variance $\sigma_x^2$.

\section{General Design Procedures}
\label{sec:overview}
	\begin{figure}[t]
		\centering
		\includegraphics[width=0.8\textwidth]{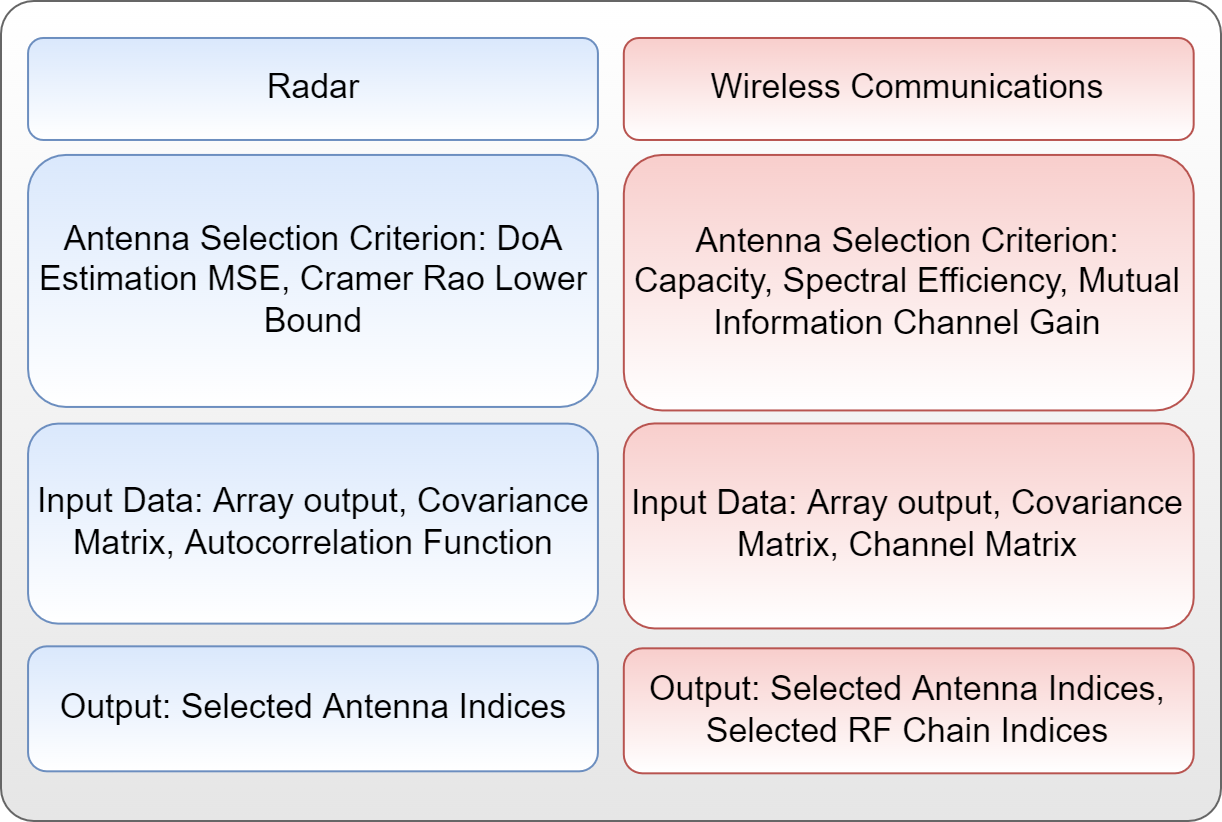} 
		\caption{Design procedures for DL-based antenna selection for the applications in radar and wireless communications. }
		\label{figAntennaSelectionDesign_radar_Comm}
	\end{figure}
	

 DL-based design for sparse arrays comprise diverse procedures that differ in performance criteria, performance metrics and input/output data types for the learning model (Fig.~\ref{figAntennaSelectionDesign_radar_Comm}). For instance, in wireless communications, sparse arrays manifest as antenna selection problem to reduce the complexity of common communications tasks such as channel estimation, beamforming, resource allocation, and user localization~\cite{antennaSelectionForMIMO,antennaSelectionMulticasting}. On the other hand, other sparse geometries (e.g., coprime and nested arrays) are used to improve the DoF of the overall system, especially in radar applications~\cite{coprime_PPV_Vaidyanathan2010Oct,nestedArray,superNestedArray,genCoprime_Qin2015Jan}. 

    The data-driven nature of DL allows us to employ a ground-truth dataset to find the best sparse array configuration. While some methods use DL, others choose simpler machine learning methods with shallow learning models, e.g. simple feed-forward neural networks~\cite{deepLearningScience,mimoDeepLearningPHY}. Although defining the optimum sparse array search as a binary segmentation problem is theoretically possible, current DL-based models do not consider the problem as a multilabel classification task.  We summarize some major approaches below.\\\\
        \noindent\textbf{Antenna Selection Setups} The potential array setups are translated into labels and a classification model is trained to find the optimum array structure. If the array setups are required to be in a specific configuration, e.g., minimum hole array (MHA), MRA, or nested arrays, either a constraint is added to the loss function to guarantee a desired architecture, or the set of possible sparse array designs is reduced to a set of sparse arrays following further constraints by a searching algorithm beforehand. The learned sparse array can be tested with any of the DoA estimation methods. In \cite{elbir2019cognitive}, a convolutional neural network (CNN) is used on a dataset composed of an $N\times N\times 3$ matrix representing real, imaginary, and angle components of a complex covariance matrix as input and each possible subarray for a given ULA as a class label as output. The problem is defined as the minimization of the Cram\'er-Rao bound (CRB), which limits the minimum variance of the correlation matrix in the DoA estimation problem. The learning model returns an appropriate subarray for individual received data, and thus works as a cognitive antenna array. The model is intended to be used as a cognitive selector, meaning that the model is used in real-time to determine the best sparse array to use in analysis by reading the incoming signal and changing the selected sparse array at a given period.
	
    The model in \cite{elbir2019cognitive} returns suboptimal 1-D arrays. The final arrays include elements at the end of the ULA grid without including surrounding elements. This issue results in coarray having many holes, causing high peak sidelobes and reducing the performance. To tackle this problem,  \cite{df_arrayDesign_Wandale2021Apr} restricted the available antenna subarrays to the ones with minimal hole setup. Here, $k$-essential property is used to eliminate sensors from a ULA, which shows whether the coarray changes from the ULA or not when the elements are removed. The $M$-$N$ sensors that are not essential can thus be discarded. The results show that the estimated subarrays provide a better beampattern with suppressed peak side lobes and a narrower main lobe. Estimated sparse arrays of this model also show much lower root mean-squared error  (RMSE) values compared to \cite{elbir2019cognitive}. 
    
     Among model-based learning approaches, \cite{df_antennaSelection_online_Aboutanios2021Jun} performed antenna selection from a CRB minimization problem defined as a weight vector selection that gives the minimum CRB for a given  number of non-zero weights at the output of the learning model. To make the original problem tractable, Dinkelbach approximation was used and solved by applying \textit{ad}v\textit{a}nced \textit{grad}ient descent (AdaGrad) algorithm on a loss function. 
    \\\\	
    \noindent\textbf{DoA Estimation Setups} Some DL-based DoA estimation models include sparsity of the array design in their architecture in order to exploit the advantages of the sparse arrays. Model-based methods require the covariance matrix to be full, which is impossible in sparse array structures. DL-based methods specifically designed to work with sparse arrays are proven to be powerful enough to overcome this issue and learn an accurate mapping between sparse covariance matrices and arrival angles~\cite{elbir2019cognitive,elbirSampTA}. These models can also utilize the coarray principle and detect more signal sources than the number of sensors in the sparse array.
	
	In \cite{df_dist_sparse_array_Pavel2021Oct}, a two-layer CNN was trained to estimate DoA angles. The array structure is a combination of different subarrays with the same number of elements. The position between different subarrays is assumed to be unknown while only the relative position of elements in each subarray is known. The model input is the fusion of the estimated covariance matrix of each subarray created with an element-wise sum and vectorization while the output yields the DoA angles. It is shown that the learning model can learn to detect more signal sources than the number of elements thanks to employing sparse arrays.
	
	A deep CNN model called RFDOA-Net was designed in \cite{df_UAV_Akter2021Sep} to estimate the DoA angles for unmanned aerial vehicles (UAVs) with a complex architecture using multi-scale feature processing. The learning model was trained with a synthetic dataset with different signal-to-noise-ratio (SNR) values over a 5-element sparse coarray and used raw antenna inputs converted into an input matrix of antenna elements with respect to snapshots. The output data was a vector of the probability distribution of the DoA angles.
 
	In \cite{df_featureEngineeringKulkarni2021Oct}, CNN-based DoA estimation was studied with nested arrays. The input of the CNN was a proxy spectrum obtained from the autocorrelation of the nested array output. The output layer of the CNN included the angular spectra, of which the highest peaks corresponds to the source directions~\cite{elbir_DL_MUSIC}. 
	
	Some models are designed to capture both azimuth and elevation angles with a 3-D sparse array structure. In \cite{df_coprimeMic_Gohil2021Jul},  CNN was applied on the 2-D readings of a coprime circular conformal (stacked array arrangement) microphone array (C3MA) to estimate the DoA angles of audio signals. The model took the cross-correlation value of individual microphones as input while the output yields the 2-D DoA angles. 
    
\section{Cognitive Sparse Array Design for DoA Estimation}
\label{sec:ant_sel}
We now examine a DL-based antenna selection procedure in detail for a cognitive radar application. The DL techniques directly fit this setting because the antenna selection problem can be considered as a classification problem, wherein each subarray designates a class. We leverage DL to consider a relatively large scale of the selection problem wherein the feature maps can be extracted to train the network for different array geometries. 
Unlike random array thinning where a fixed subarray is used for all scans, we select a new subarray based on the received data. In contrast to \cite{antennaSelectionKnapsack,sparsityEnforcingSS}, we also assume that the target DoA angle is unknown while choosing the array elements. 
    
     We employ a CNN, whose input data are the covariance samples of the received array signal. 
     The CNN architecture models the selection of $K$ best antennas out of $M$ as a classification problem wherein each class denotes an antenna subarray. In order to create the training data, we choose those subarrays which estimate DoA with the lowest minimal bound on the mean-squared-error (MSE) so that the selected subarray can yield the best performance for the given scenario. We consider minimization of CRB as the performance benchmark in generating training sets for 1-D ULA and 2-D geometries such as uniform circular arrays (UCA) and randomly deployed arrays (RDA). For ULAs, the network is also trained with data obtained by minimizing Bayesian bounds such as the Bobrovsky-Zakai bound (BZB) and Weiss-Weinstein bound (WWB) on DoA \cite{performanceBoundsWWB} because these bounds provide better estimates of MSE at low SNRs. In particular, BZB-based selection has been shown to have the ability to control the sidelobes and avoid ambiguity in DoA estimation \cite{antennaSelectionCognitive,antennaSelectionCognitive2}.

\subsection{Signal Model}
	Consider a phased array antenna with $M$ elements where each element transmits a pulsed waveform $s(t_i)$ towards a Swerling Case 1 point target for $i = 1, \cdots, T$, where $T$ is the number of data snapshots. Since we are interested only in DoA recovery, the range and Doppler measurements are not considered and target's complex reflectivity is set to unity. {Assumption of a Swerling I model implies that the target parameters remain constant for the duration of the scan}. We characterize the target through its DoA $\Theta=(\theta,\phi)$ where $\theta$ and $\phi$ denote, respectively, the elevation and the azimuth angles with respect to the radar. The radar's pulse repetition interval and operating wavelength are, respectively, $T_s$ and $\lambda= c_0/f_0$, where $c_0 = 3\times10^8$ ms$^{-1}$ is the speed of light and $f_0=\omega_0/2\pi$ is the carrier frequency. 

 To further simplify the geometries, we suppose that the targets are far enough from the radar so that the received signal wavefronts are effectively planar over the array. The array receives a narrowband signal reflected from a target located in the far-field of the array at $\Theta$. Then, the output of the sensor array is \textcolor{black}{\cite{crbStoicaNehorai}}
	\begin{align}
	\label{signalModel}
	\mathbf{y}(t_i) = \mathbf{a}(\Theta)s(t_i) + \mathbf{n}(t_i),\hspace{10pt} 1\leq i\leq T,
	\end{align}
	where $T$ is the number of snapshots, $\mathbf{y}(t_i) = [y_1(t_i),\dots, y_M(t_i)]^T$ and ${y}_m(t_i)$ denotes the output of the $m$-th sensor for the $i$-th snapshot, $\mathbf{n}(t_i)$ $ = [n_1(t_i),\dots, $ $n_M(t_i)]^T$ is the noise vector and $n_m(t_i)$ is zero-mean spatially and temporarily white Gaussian noise with variance $\sigma_n^2$,  $\mathbf{a}(\Theta) = [a_1(\Theta),\dots, a_M(\Theta)]^T$ is the $M\times 1$ steering vector. The $m$-th element of $\mathbf{a}(\Theta)$ is
	\begin{align}
	a_m(\Theta) = \exp\left\{-j\frac{2\pi}{\lambda}\mathbf{p}_m^T\mathbf{r}(\Theta)\right\},
	\end{align}
	where $\mathbf{r}(\Theta)$ depends on the source direction as
	\begin{align}
	\mathbf{r}(\Theta) = [\cos(\phi)\sin(\theta), \sin(\phi)\sin(\theta), \cos(\theta)]^T,
	\end{align}
	and $\mathbf{p}_m = [x_m,y_m,z_m]^T$ is the position of the $m$-th sensor in the Cartesian coordinate system. Expanding the inner product $\textbf{p}_m^T \textbf{r}(\Theta)$ in the array steering vector gives $a_m(\Theta) = \exp\left\{{-}j\dfrac{2\pi}{\lambda} (p_{x_m} \mu + p_{y_m} \nu + p_{z_m}\xi)\right\}$, where $\mu = \cos(\phi)\sin(\theta)$, $\nu = \sin(\phi)\sin(\theta)$, and $\xi=\cos(\theta)$. Evidently, $a_m(\Theta)$ is a multi-dimensional harmonic. Once the frequencies $\mu$, $\nu$, and $\xi$ in different directions are estimated, the DoA angles are obtained using the relations 
\begin{align}
\label{eq:thetaphiEst}
\theta = \tan^{-1}\left(\frac{\nu}{\mu}\right),\;\;\; \phi=\cos^{-1}\left(\frac{\xi}{\sqrt{\mu^2+\nu^2+\xi^2}}\right),
\end{align}
with the usual ambiguity in $[0, 2\pi]$. In case of a linear array, there is only one parameter in the steering vector whereas two parameters are involved in planar and three-dimensional arrays.
	
	In the context of sparse array selection, our goal is to choose the ``best" $K$ sensors in an $M$-element array in the sense that the lowest statistical mean-square-error (MSE), i.e., the CRB is achieved \cite{crbStoicaNehorai,friedlander}. Overall, $C = \left( \begin{array}{c} M \\ K \end{array}\right)  = \frac{M!}{K! (M-K)!}$ possible subarray choices are available. Therefore, we can treat sensor selection as a classification problem with $C$ classes. It seems impractical to visit all possible subarray configurations to arrive at the best subarray candidate. However, it has been shown \cite{elbir2019cognitive,elbirQuantizedCNN2019,elbirSampTA} that many subarray candidates yield the same CRB level because of the non-unique placement of sensors within the array. Hence, the distinct number of subarrays is very small, which makes this technique very practical. 
	Note that the literature suggests other statistical bounds \cite{performanceBoundsWWB} for DoA estimation but a closed-form solution of only CRB is available for higher dimensional arrays.
	
	We observe $\Theta$ as the inner product $\mathbf{p}_m^T\mathbf{r}(\Theta)$. The exponential form of $a_m(\Theta)$ suggests that this is a multi-dimensional spatial harmonic whose frequencies (and hence, DoAs) can be extracted through conventional as well as sparse reconstruction algorithms \cite{elbir2019cognitive,mishra2017high}. The uniqueness of spatial harmonic retrieval \cite{nion2010tensor} is directly related to the number of sensors in the array. For a URA of size $M_1\times M_2$, at least $M_1M_2 - \text{min}(M_1,M_2)$ sensors are required for a perfect DoA retrieval in a noiseless setting. Hence, in any sparse sensor array selection, $K$ must satisfy these guarantees.

 In general, the target's position changes little during consecutive scans while a phased array can switch very fast from one antenna configuration to the other. {Here, we consider the following scan strategy for the radar: at the beginning (the very first scan), all $M$ antennas are active and the received signal from this scan is fed to the network. Our goal is to find an optimal antenna array for the next scan in which only $K$ antennas will be used. The radar continues to use this subarray for a few subsequent scans. After surveying the target scene with this optimal subarray for a predetermined number of scans, the radar switches back to the full array for a single scan. The received signal from this full array scan is then used to find a new, optimal subarray for the subsequent few scans. The frequency of choosing a new subarray can be decided off-line based on the nature of the target and analysis of previous observations.} This switching of elements between different scans is a cognitive operation because a new array is determined in every few scans based on received echoes from the target scene. 

	\begin{figure}[t]
		\centering
		{\includegraphics[width=\textwidth]{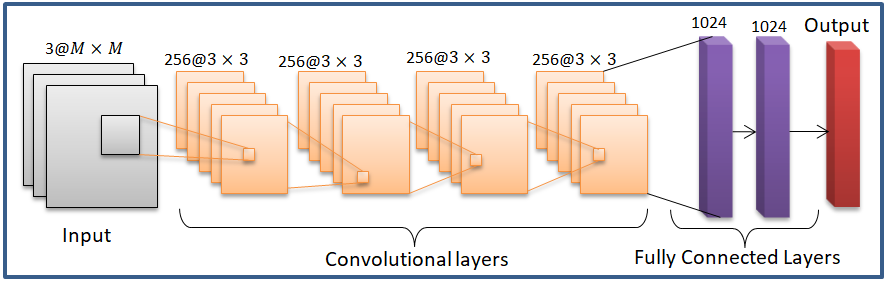}}
		\caption{Structure of the CNN for sensor selection \cite{elbir2020TL}.}
		\label{figNetwork}
	\end{figure}

\subsection{Antenna Selection via Deep Learning}
A DL network (see Fig.~\ref{figNetwork}) is defined as a non-linear mapping which categorizes and clusters the input data. Let $\mathcal{D} = \{\mathcal{D}^{(1)},\dots, \mathcal{D}^{(\textsf{D})}  \}$ and $\mathcal{Y} = \{\mathcal{Y}^{(1)},\dots, \mathcal{Y}^{(\textsf{Y})}\}$ denote the input and output labels for a dataset. Then, the deep classification network is represented as $\Sigma (\mathcal{D}) = \mathcal{Y}$ mapping the input data to the output labels which represent the best subarray indices. 

Assume that an antenna subarray composed of $K$ antennas is to be selected from an $M$-element antenna array. There are $C = \left( \begin{array}{c} M \\ K \end{array}\right) $ possible choices. This can be viewed as a classification problem with $C$ classes each of which represents a different subarray. Let $\mathcal{P}_k^c = \{p_{x_k}^c,p_{y_k}^c,p_{z_k}^c\}$, $k=1,\dots,K$, be the set of the $k$th antenna coordinates in the $c$th subarray. Then, the $c$th class consisting of the positions of all elements in the $c$th subarray is
	\begin{align}
	\mathcal{S}_c =  \{\mathcal{P}_1^c,\dots,\mathcal{P}_K^c  \},
	\end{align}
    and all classes are given by the set
	\begin{align}
	\mathcal{S} = \{\mathcal{S}_1,\mathcal{S}_2,\dots, \mathcal{S}_C  \}.
	\end{align}
 Next, we design a CNN to solve this classification problem by selecting the positions of the antenna subarray that provides the best DoA estimation performance. Note that, for an operational radar, generation of an artificial or simulated dataset to train the DL network is not necessary. Instead, the network can train itself with the data acquired by the radar during previous scans. During the test phase, the DoA angles are unknown to the network. The CNN accepts the features from the estimated covariance matrix and outputs a new array. This stage is, therefore, cognitive because the radar is adapting the antenna array in response to the received signal. In the following, we present the details of input and output design of the deep network.\\\\
	\noindent\textbf{Input Data} 
	The input to the DL network are the covariance matrices of the received signal. In particular, we use the real, imaginary and the phase information of the covariance matrix. Let $\mathbf{X}$ be an $M\times M\times 3$ real-valued matrix with $3$ ``channel". Hence, we have $\mathcal{D}^{(i)} = \mathbf{X}$ for $i$-th input instant. Specifically, we define the $(i,j)$-th entry of the first and the second "channel" of the input data as $[\mathbf{X}_{(:,:,1)}]_{i,j} = \operatorname{Re}\{ [\mathbf{R}]_{i,j} \}$ and $[\mathbf{X}_{(:,:,2)}]_{i,j} = \operatorname{Im}\{ [\mathbf{R}]_{i,j} \}$, respectively, where $\mathbf{R}$ is the array covariance matrix as
\begin{align}
    \mathbf{R}= \frac{1}{T}\sum_{i=1}^{T}\mathbf{y}(t_i)\mathbf{y}^H(t_i).
\end{align}
 Similarly, the third "channel" is given by $[\mathbf{X}_{(:,:,3)}]_{i,j} = \angle\{ [\mathbf{R}]_{i,j} \}$. Although real and imaginary inputs are sufficient to describe the complex covariance matrix, feeding a third quantity such as phase (or magnitude) lets the network know that the first two inputs are related to each other.\\\\
	\noindent\textbf{Labeling}
	We treat the sensor selection problem as a classification problem with $C$ classes. The class label comprises the positions of the sensor subarray corresponding to that class. Let $\mathcal{P}_c^{(k)} =  \{{x_k}^{(c)},{y_k}^{(c)},{z_k}^{(c)}  \}$ be the set of sensor coordinates in the $c$-th subarray for $k = 1,\dots,K$. Then the positions of the sensors for the $c$-th class form the set $\mathcal{Y}_c =\{\mathcal{P}_c^{(1)},\dots,\mathcal{P}_c^{(K)} \}$. Therefore, the set of all classes is $\mathcal{Y} = \{\mathcal{Y}_1,\mathcal{Y}_2,\dots,\mathcal{Y}_C\}$.
	
	In order to select the best subarrays in $\mathcal{Y}$, we compute the CRB for each element of $\mathcal{Y}$ as $c = 1,\dots,C$. Consider the $K\times 1$ subarray output
	\begin{align}
	\label{subarrayOutput}
	\mathbf{y}_c(t_i) =  \mathbf{a}_c(\Theta )s(t_i) + \mathbf{n}_c(t_i),
	\end{align}
	where $\mathbf{a}_c(\Theta)\in \mathbb{C}^{K}$ denotes the array steering vector corresponding to the subarray with position set $\mathcal{Y}_c$. Let $\mathbf{R}_c= \frac{1}{T}\sum_{i=1}^{T}\mathbf{y}_c(t_i)\mathbf{y}_c^H(t_i)$ be the $K\times K$ subarray sample covariance matrix for the $K\times 1$ subarray output $\mathbf{y}_c(t_i)$. We denote the partial derivatives of $\mathbf{a}_c(\Theta)$ with respect to $\theta$ and $\phi$ by $\dot{\mathbf{a}}_c(\theta) = \frac{\partial \mathbf{a}_c(\Theta)}{\partial_{\theta}}$ and  $\dot{\mathbf{a}}_{c}(\phi) = \frac{\partial \mathbf{a}_c(\Theta)}{\partial_{\phi}}$, respectively. The signal and noise variances are $\sigma_s^2$ and $\sigma_n^2 $, respectively.
	
	The CRBs for $\theta$ and $\phi$ in a single source scenario are \cite{crbStoicaNehorai}
	\begin{align}
	\kappa(\theta,\mathcal{Y}_c) = \frac{\sigma_n^2}{2T\operatorname{\mathbb{R}e}\bigg\{ \boldsymbol{\Pi}_{\theta} \odot (\sigma_s^4\mathbf{a}_c^H(\Theta)\mathbf{R}_c^{-1} \mathbf{a}_c(\Theta))\bigg\}},\label{CRB_Theta}\\
	\kappa(\phi,\mathcal{Y}_c) = \frac{\sigma_n^2}{2T\operatorname{\mathbb{R}e}\bigg\{ \boldsymbol{\Pi}_{\phi}\odot (\sigma_s^4\mathbf{a}_c^H(\Theta)\mathbf{R}_c^{-1} \mathbf{a}_c(\Theta))\bigg\}},\label{CRB_Phi}
	\end{align}
	where
	\begin{align}
	\boldsymbol{\Pi}_{\theta}  = \dot{\mathbf{a}}_{c}^H(\theta) \left[\mathbf{I}_K - \frac{\mathbf{a}_c(\Theta) \mathbf{a}_c^H(\Theta)}{K} \right]\dot{\mathbf{a}}_{c}(\phi),\\
	\boldsymbol{\Pi}_{\phi} = \dot{\mathbf{a}}_{c}^H(\phi) \left[\mathbf{I}_K - \frac{\mathbf{a}_c(\Theta) \mathbf{a}_c^H(\Theta)}{K} \right]\dot{\mathbf{a}}_c(\theta).
	\end{align}
	We define the absolute CRB \cite{ye2008two} for the directions $\Theta$ and $\mathcal{Y}_c$ as the root-mean-square (RMS) value 
	\begin{align}
	\label{computeCRB}
	\kappa(\Theta,\mathcal{Y}_c) = \frac{1}{\sqrt{2}}[\kappa(\theta,\mathcal{Y}_c)^2 +\kappa(\phi,\mathcal{Y}_c)^2]^{1/2}.
	\end{align}
	For simplicity, we select $\sigma_s^2=1$ and define the signal to noise ratio in the training data as SNR$_{\text{TRAIN}} = 10\log_{10}(\sigma_s^2/\sigma_n^2)$.

	Once $\kappa(\Theta,\mathcal{Y}_c)$ is computed for $c = 1,\dots,C$, the best subarray label $\mathcal{B}_{\bar{c}}$ is
	\begin{align}
	\label{ReducedSet}
	\mathcal{B}_{\bar{c}} = \arg \min_{c = 1,\dots,C }  \kappa(\Theta,\mathcal{Y}_c).
	\end{align}

	\begin{table}[t]
		\caption{ Number of classes $C$ and the reduced number of classes $\bar{C}$ for a UCA with $M=16$.   \label{tableComparisonForNumberOfClasses}
  }
		{\begin{tabular}{|c|c|c|c|c|c|c|}
				\hline
				&$K=3$ &$K=4$ &$K=5$ &$K=6$ &$K=7$ &$K=8$  \\
				\hline
				$C$& $560$ &$1820$ &$4368$ &$8008$ &$11440$ &$12870$  \\
				$\bar{C}$&$16$ &$10$ &$16$ &$11$ &$16$ &$16$  \\
				\hline
		\end{tabular}}{}
	\end{table}

	Here, the subscript $(\cdot)_{\bar{c}}$ denotes the index of best subarrays, $\bar{c} = 1,\dots, \bar{C}$, where $\bar{C}$ is the number of best subarrays. {\color{black}As $K$ increases, $C$ becomes very large. This makes the classification operation very difficult. However, experiments reveal that most of the sensor subarrays yield the same $\kappa(\Theta,\mathcal{Y}_c)$ because the non-unique sensor positions are common in many subarray combinations. Hence, $\bar{C} \ll C$ implying that only a handful of classes yield the lowest estimation errors \cite{elbir2019cognitive,elbirQuantizedCNN2019}. In Table~\ref{tableComparisonForNumberOfClasses}, we present the comparison of $C$ and $\bar{C}$ for a UCA with $M=16$ antennas. We note that $\bar{C}$ is very small, which leads to an effective classification performance.}  After computing all best subarray indices, we finally construct the best subarray set as $\mathcal{B} = \{\mathcal{B}_{1},\dots,\mathcal{B}_{\bar{C}} \}$, where $\mathcal{B} \subset \mathcal{Y}$.

	\begin{algorithm}[t!]
		\begin{algorithmic}[1]
			\caption{Training data generation.}
			\Statex {\textbf{Input:} \label{alg:TrainingDataGen.} Sensor positions $\{\mathbf{p}_m\}_{m=1}^M$, $K$, $T$, number of data realizations $L$, number of directions $P$ and SNR$_{\text{TRAIN}}$}.
			\Statex {\textbf{Output:} Training data $\mathcal{T}$ with dimensions $\{M\times M\times3 \times LP, LP\}$.}
			\State Generate $P$ DoA angles $\Theta_p = (\theta_p,\phi_p)$ for $p =1,\dots,P$.
			\State \textbf{for} $1\leq p \leq P$ \textbf{do}
			\State \indent \textbf{for} $1\leq l \leq L$ \textbf{do}
			\State \indent Generate the array output $\{ \mathbf{y}^{(l,p)}(t_i)\}_{i=1}^T$ as
			\begin{align}
			\mathbf{y}^{(l,p)}(t_i) = \mathbf{a}(\Theta_p)s^{(l,p)}(t_i) + \mathbf{n}^{(l,p)}(t_i), \nonumber
			\end{align}
			\indent for $s^{(l,p)}(t_i)\hspace{-1pt} \sim \mathcal{CN}(0,\sigma_s^2)$, $\mathbf{n}^{(l,p)}(t_i)\hspace{-2pt} \sim\hspace{-1pt}\mathcal{CN}(0,\sigma_n^2\mathbf{I})$.
			\State \indent Construct all $K\times 1$ subarray output configurations \par \indent $\mathbf{y}_c^{(l,p)}(t_i)$ as in (\ref{subarrayOutput}) from $\mathbf{y}^{(l,p)}(t_i)$ for $c = 1,\dots,C.$
			\State \indent Compute $\kappa(\Theta_p,\mathcal{Y}_c)$ for $c = 1,\dots,C$  by using the \par \indent covariance matrices ${\mathbf{R}}_c^{(l,p)}$.
			\State \indent Using $\kappa(\Theta_p,\mathcal{Y}_c)$, find the best subarray index as
			\par \indent  $\mathcal{B}_{\bar{c}}^{(l,p)}$ from (\ref{ReducedSet}).
			\State \indent Compute the full array covariance matrix $\mathbf{R}^{(l,p)}$ 
			\par \indent from  $\mathbf{y}^{(l,p)}(t_i)$, $i = 1,\dots,T$.
			\State \indent Construct the input data $\mathbf{X}^{(l,p)}$ as
			\begin{align}
			[\mathbf{X}_{(:,:,1)}^{(l,p)}]_{i,j}& = \operatorname{Re}\{ [\mathbf{R}^{(l,p)}]_{i,j} \},\nonumber \\
			[\mathbf{X}_{(:,:,2)}^{(l,p)}]_{i,j}& = \operatorname{Im}\{ [\mathbf{R}^{(l,p)}]_{i,j} \}, \nonumber\\
			[\mathbf{X}_{(:,:,3)}^{(l,p)}]_{i,j} &= \angle\{ [\mathbf{R}^{(l,p)}]_{i,j} \}.\nonumber
			\end{align}
			\State \indent Design the output label as $z^{(l,p)} = \mathcal{B}_{\bar{c}}^{(l,p)}$.
			\State \indent \textbf{end for} $l$
			\State \textbf{end for} $p$
			\State Construct training data by concatenating the input-output pairs: \noindent \small $ \mathcal{T} = \{ (\mathbf{X}^{(1,1)}, z^{(1,1)}), (\mathbf{X}^{(1,2)}, z^{(2,1)}),\dots, $ $(\mathbf{X}^{(1,L)}, z^{(L,1)}),$ $(\mathbf{X}^{(2,1)}, z^{(1,2)})\dots, (\mathbf{X}^{(P,L)}, z^{(L,P)})\}.$ \normalsize
		\end{algorithmic}
	\end{algorithm}

 Algorithm~\ref{alg:TrainingDataGen.} lists the steps to generate the training data by incorporating the input and labels, as discussed above. The training data is then fed to the deep network represented by $\Sigma(\cdot): \mathbb{R}^{M\times M\times 3} \rightarrow \mathcal{Y}$ that maps the input data $\mathbf{X}$ to the corresponding class in $\mathcal{Y}$.
\\\\
 	\noindent\textbf{Network Architecture} 
	Figure~\ref{figNetwork} illustrates the  deep network architecture for sensor selection. For a  multi-layer network, the non-linear function $\Sigma(\cdot)$ is represented by the inner layers as
	\begin{align}
	\Sigma(\mathcal{D}) = f^{(15)} \big(f^{(14)} ( \dots f^{(2)}(f^{(1)}( \mathcal{D} )   )   )   \big)  = \mathcal{Y},
	\end{align}
	where the first layer $f^{(1)}$ is the input layer and $f^{(i)}_{i \in\{2,4,6,8\}}$ denote the convolutional layers, each of which has 256 filters of size $3\times 3$. 	The arithmetic operation of a single filter of a  convolutional layer is defined for an arbitrary input  $\bar{\bf X} \in \mathbb{R}^{d_{x}\times d_{x}\times V_x}$ and output $\bar{\bf Y} \in \mathbb{R}^{d_{y}\times d_y\times V_{y}}$ as 
	\begin{align}
	\bar{\bf Y}_{p_y,v_y} = \sum_{p_k,p_x} \langle \bar{\bf W}_{v_y,p_k}, \bar{\bf X}_{p_x} \rangle,
	\end{align}
	where   $d_x \times d_y$ is the size of the convolutional kernel, $V_x \times V_y$ is the size of the response of a convolutional layer, $\bar{\bf W}_{v_y,v_k}\in \mathbb{R}^{V_x}$ denotes the weights of the $v_y$-th convolutional kernel, and $\bar{\bf X}_{p_x} \in \mathbb{R}^{V_x}$ is the input feature map at spatial position $p_x$. Hence, we define $p_x$ and $p_k$ as the two-dimensional (2-D) spatial positions in the feature maps and convolutional kernels, respectively \cite{quantizedCNN_Unified}.
	
	The $10$-th and $12$-th layer are fully connected with 1024 units whose $50\%$ is randomly selected during training to avoid overfitting. A fully connected layer maps an arbitrary input $\bar{\bf x}\in \mathbb{R}^{U_x}$ to the  output $\bar{\bf y}\in \mathbb{R}^{U_y}$ by using the weights  $\bar{\bf W} \in \mathbb{R}^{U_{x}\times U_{y}}$. Then, the $u_y$-th element of the output of the layer is the inner product
	\begin{align}
	\bar{\bf y}_{u_y} = \langle \bar{\bf W}_{u_y}, \bar{\bf x} \rangle = \sum_{i} {[\bar{\bf W}}]_{u_y,i}^\textsf{T} \bar{ \bf x}_i ,
	\end{align}
	for $u_y = 1,\dots, U_y$ and  $\bar{\bf W}_{u_y}$ is the $u_y$-th column vector of $\bar{\bf W}$, and $U_x = U_y = 1024$ is selected for $f^{(14)}$.
	
	After each convolutional and fully connected layers (i.e., $f^{(i)}_{i\in \{3,5,7,9,11,13\}}$), there is a rectified linear unit ($\mathrm{ReLU}$) layer  where $\mathrm{ReLU}(x) = \max(0,x)$. The $\mathrm{ReLU}$ layers are powerful in constructing the non-linearity of the deep network as well as providing non-negative output at the output layers, which is very useful for classification networks. The $14$-th layer has a classification layer with $\bar{C}$ units, where a $\mathrm{softmax}$ function is used to obtain the probability distribution of the classes. The  $\mathrm{softmax}$ layer is defined for an arbitrary input $\bar{\mathbf{x}}\in \mathbb{R}^{D}$ as $\mathrm{softmax}(\bar{{x}}_i) = \frac{\exp \{\bar{x}_i\} }{\sum_{i=1}^{D} \exp \{\bar{x}_i\} }$. The last layer $f^{(15)}$ is the classification layer.


	\begin{figure*}[t]
		\centering
            \subfloat[][]{\includegraphics[width=.3\textwidth]{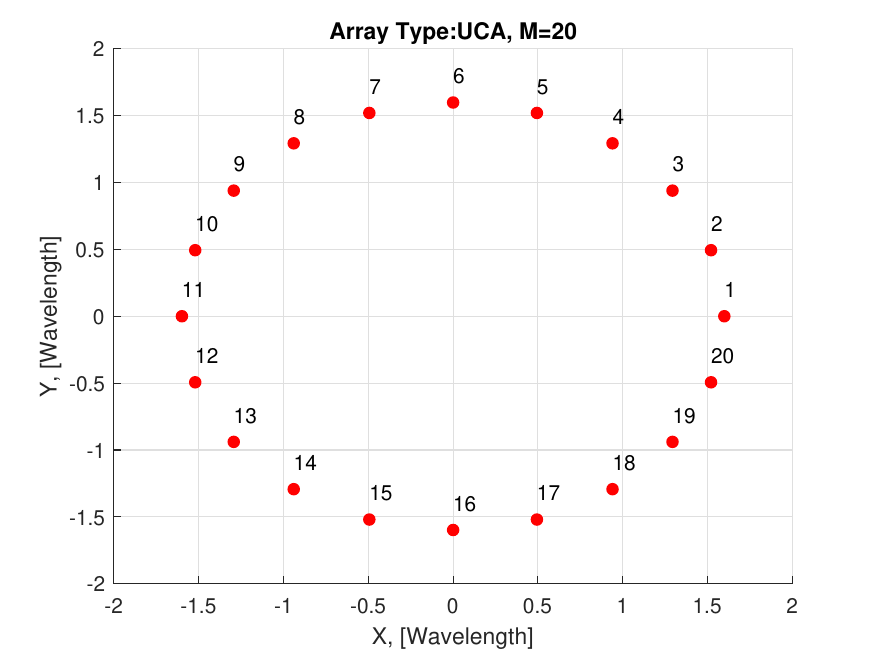}%
			}
		\subfloat[][]{\includegraphics[width=.3\textwidth]{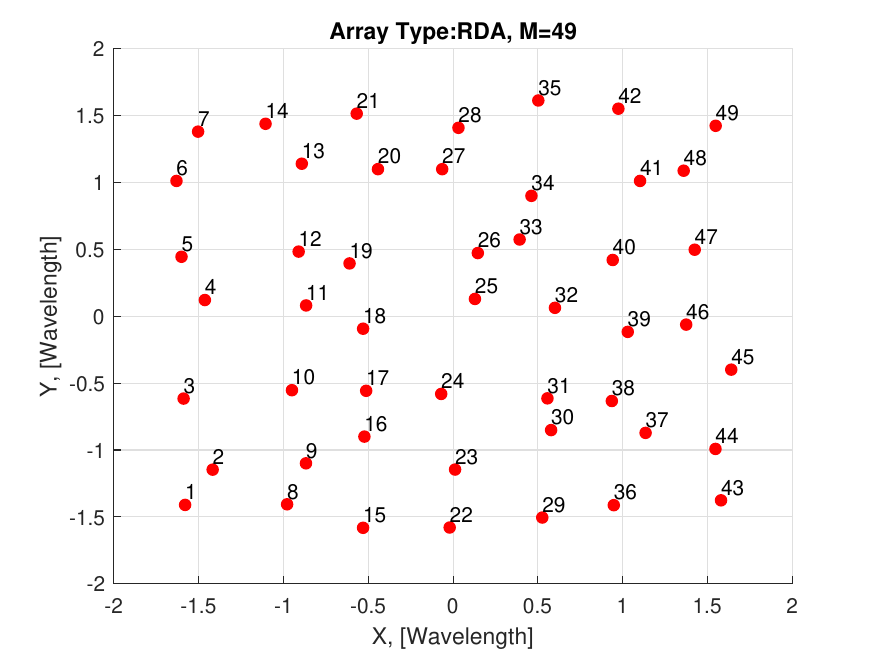}%
			 }
		\subfloat[][]{\includegraphics[width=.3\textwidth]{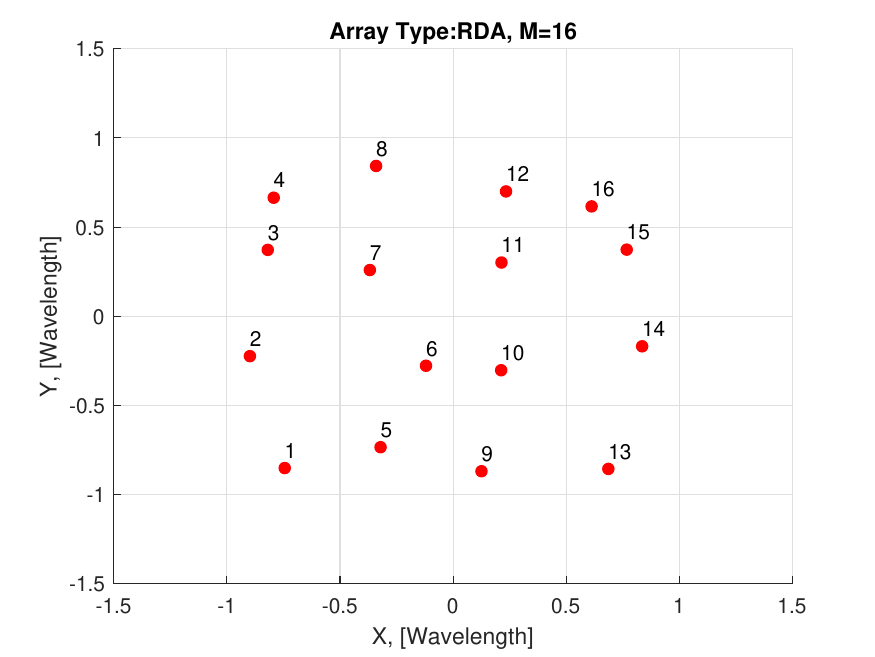}%
		}
		\caption{Placement of antennas for a) UCA with $M=20$ elements, b) RDA with $M=49$ and c) RDA with $M=16$ \cite{elbir2019cognitive}.}
		\label{figArrayPlacement1D}
	\end{figure*}

\subsection{Numerical Experiments}
 We now investigate antenna selection performance of various array geometries such as UCAs and RDAs as shown in Fig. \ref{figArrayPlacement1D}. The classification accuracies of training and validation data for UCA and RDA are listed in Table \ref{tableAccuracies}. We can see that the learning model achieves high accuracy when $M$ is large and SNR$_{\text{TEST}}\geq 10$dB.

	\begin{table}[ht]
		\caption{The accuracy percentages for training and validation datasets in 1-D and 2-D scenario. \label{tableAccuracies}}
		{\begin{tabular}{c|cc|cc}
			\hline
			&\multicolumn{2}{|c|}{1-D, UCA with $M=20$, $K=6$.} &\multicolumn{2}{c}{1-D, UCA with $M=45$, $K=5$.} \\
			\hline
			SNR$_{\text{TRAIN}}$ & Training & Validation & Training  & Validation\\
			10 dB &65.2\%  & 68.7\% & 98.7\% &97.8\%  \\
			15 dB &98.1\%  & 98.5\% & 99.0\% &98.7\%  \\
			20 dB & 99.2\%& 99.5\%  & 100\% &99.7\%  \\
			25 dB & 99.4\% & 99.8\%  & 100\% &100\%  \\
			30 dB & 100\% & 100\%  &100\%  &100\%   \\
			inf dB & 100\% & 100\%  & 100\% &100\%  \\
   \hline
     & \multicolumn{2}{|c}{1-D, RDA with $M=49$, $K=5$.} &\multicolumn{2}{|c}{2-D, RDA with $M=16$, $K=6$.}\\
			SNR$_{\text{TRAIN}}$ & Training & Validation & Training  & Validation\\
   10 dB &97.8\% & 95.7\% &8.1\% & 10.7\%\\
 15 dB   &99.9\%  &  98.6\%&60.1\%  &  63.2\%\\
			20 dB    & 97.5\% & 98.1\% &80.8\% & 80.6\%\\
			25 dB    & 100\% & 100\% &88.9\% & 89.2\%\\
			30 dB   & 100\% & 100\% & 82.6\% & 83.2\% \\
			inf dB  &100\%  & 100\% &85.0\%  & 83.9\%\\
			\hline
		\end{tabular}}
		
	\end{table}

\begin{figure}[ht]
		\centering
		\subfloat[][]{\includegraphics[width=.31\textheight]{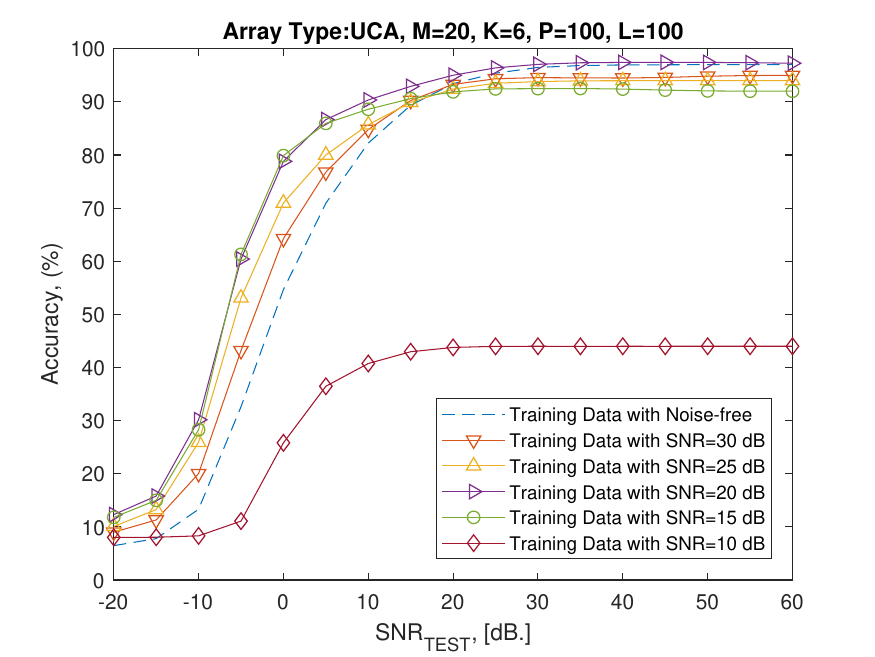}}
        \subfloat[][]{\includegraphics[width=.31\textheight]{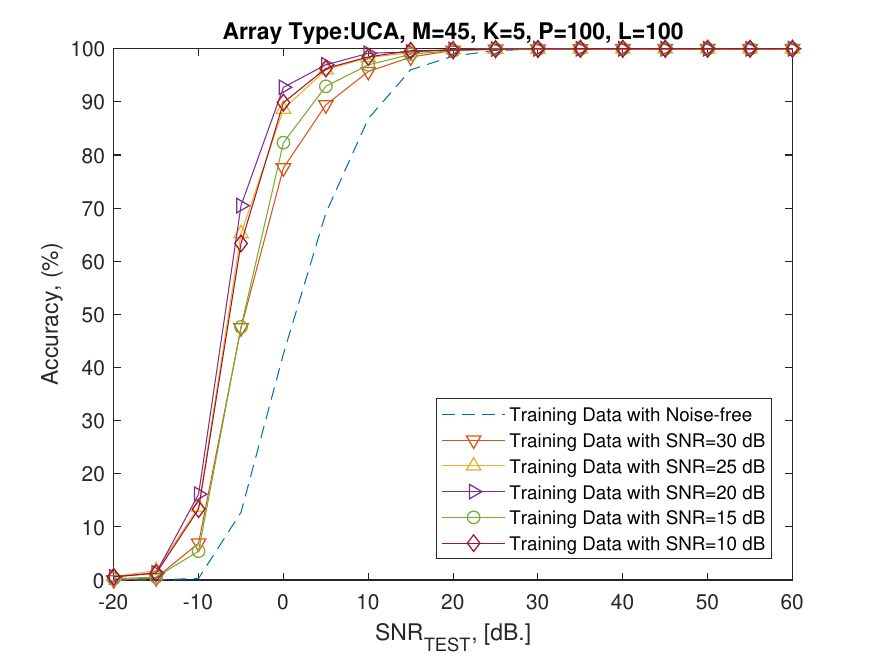}}
		\caption{{Performance of test dataset using CNN with respect to SNR$_{\text{TEST}}$ for a UCA with (a) $M=20$, $K=6$, and (b) $M=45$, $K=5$, respectively \cite{elbir2019cognitive}. }}
		\label{figSNRTest1DUCAM20}
	\end{figure}

Figure~\ref{figSNRTest1DUCAM20} shows the classification performance of the CNN for $J_{\text{TEST}}=100$ Monte Carlo trials.  Figure~\ref{figSNRTest1DUCAM20} also shows the performance of the noisy test data when the network is trained with noise-free dataset; it's performance degrades especially at low SNR levels. These observations imply that noisy training datasets should be used for robust classification performance with the test data. On the other hand, when the training data is corrupted with strong noise content (e.g., SNR$_{\text{TRAIN}}\leq 10$dB), then despite using the noisy training data, the  CNN model does not recover from poor performance at low SNR$_{\text{TEST}}$ regimes. The performance at low SNRs can be improved when the size of the array increases and, as a result, the input data is huge and the SNR is enhanced due to large $M$. As an example, Fig. \ref{figSNRTest1DUCAM20}b illustrates the performance of the network for UCA with $M=45$ and $K=5$, where the network provides high accuracy for a wide range of SNR$_{\text{TEST}}$ compared to the scenario in Fig. \ref{figSNRTest1DUCAM20}a.

	\begin{figure}[ht]
		\centering
		\subfloat[][]{\includegraphics[width=.31\textheight]{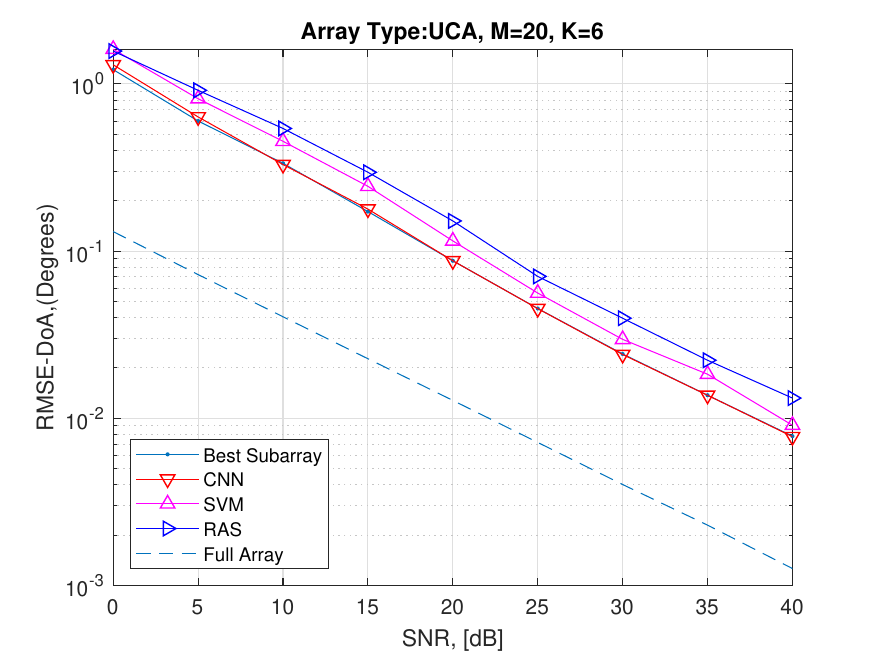}}
        \subfloat[][]{\includegraphics[width=.31\textheight]{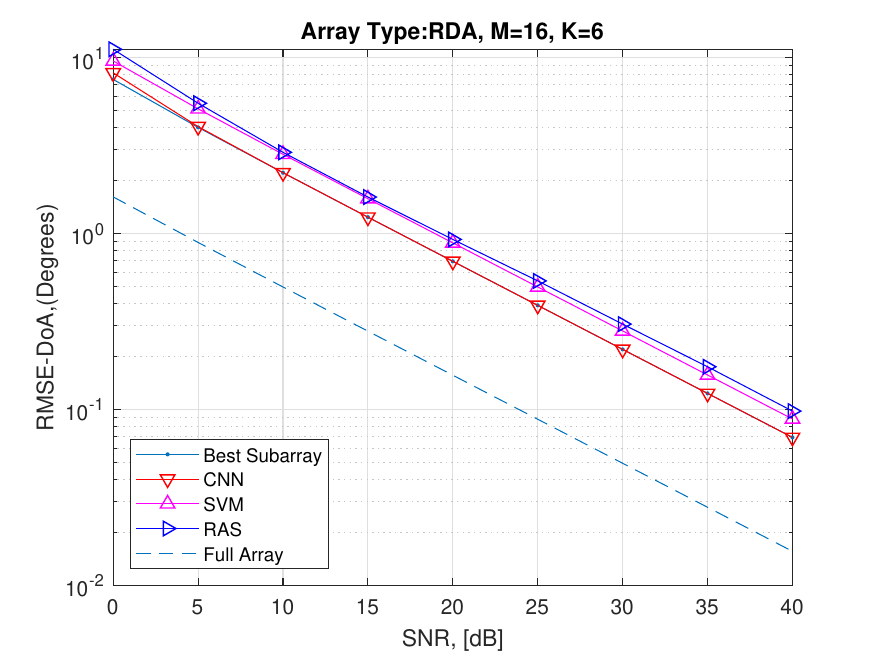}}
		\caption{DoA estimation performance with respect to SNR. SNR$_{\text{TRAIN}}=20$ dB. The antenna geometry is (a) UCA with $M=20$ and $K=6$, (b) RDA with $M=16$ and $K=6$ \cite{elbir2019cognitive}. }
		\label{figDOAEstimationSNRTestUCA}
	\end{figure}

	\begin{figure}[ht]
		\centering
		\subfloat[][]{\includegraphics[width=.45\textwidth]{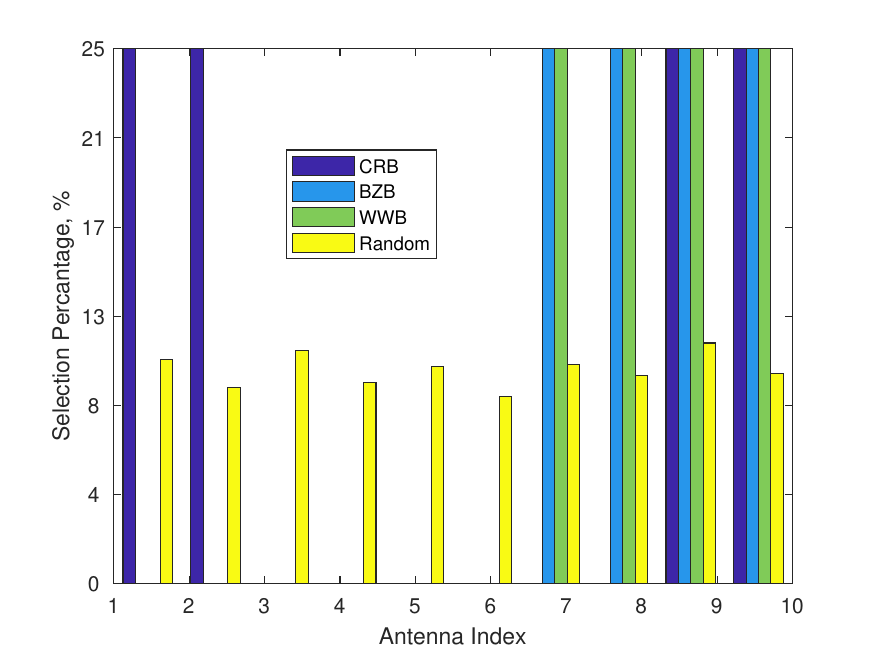}}
			\subfloat[][]{\includegraphics[width=.45\textwidth]{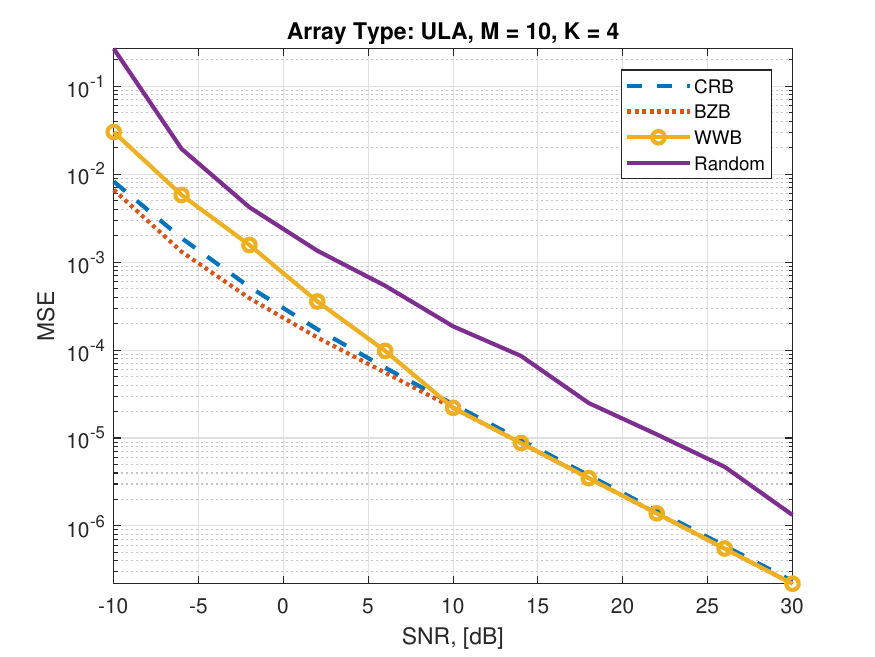}}
		\caption{(a) Antenna selection percentage over $J_{\text{TEST}}=10000$ trials. (b) MSE of DoA for selected subarrays. The array geometry is a ULA with $M=10$ and $K=4$ \cite{elbir2019cognitive}.}
		\label{figULABoundAnalysis}
	\end{figure}
DoA estimation performance of the DL-based scheme is also presented. The CNN approach is compared with support vector machine (SVM) and random antenna selection (RAS) algorithms. The selected antenna subarrays from CNN and SVM are inserted to the beamforming technique \cite{beamformingFundamentals} for DoA estimation. As a traditional technique, we consider the RAS algorithm where, instead of all subarray candidates, a number of subarray geometries are realized randomly (i.e., 1000 realizations) and their beamforming spectra is obtained by a search algorithm \cite{randomArraySelection1}. We also added the full array performance where $M=K$ for comparison. In Fig.~\ref{figDOAEstimationSNRTestUCA}a, the results are given for a UCA with $M=20$ and $K=6$ antennas to be selected. Here, "best subarray" denotes the beamforming performance of the subarray that gives the lowest CRB. It can be seen that CNN provides better performance as compared to SVM (32\% more accurate) and RAS (72\%) and it approaches the performance of the "best subarray" as expected from the accuracy results given in Fig.~\ref{figSNRTest1DUCAM20}a. SVM performs poorer due to its lower antenna selection accuracy. We present 2-D DoA estimation results in Fig.~\ref{figDOAEstimationSNRTestUCA}b for RDA with $M=16$ and $K=6$. Similar observations are obtained for the 2-D case as compared to the 1-D scenario.
     
We further compare the DoA estimation performance of the selected subarrays with full array ($M=K$) performance in both Fig.~\ref{figDOAEstimationSNRTestUCA}a and Fig.~\ref{figDOAEstimationSNRTestUCA}b. While there is a gap between subarray and the full array performances, antenna selection provides less computation and cost.

We also analyze the effect of the performance metrics on the antenna selection and DoA estimation accuracy by employing the simplest and most common geometry of a ULA. For creating the training data, we employed three bounds: CRB, BZB and WWB \cite{performanceBoundsWWB}. The network was trained for $M=10$, $K=4$, $L_{\text{TRAIN}}=100$ snapshots, $T_{\text{TRAIN}}=100$ signal and noise realizations, and $P_{\text{TRAIN}}=100$ DoA angles. The number of uniformly spaced azimuthal grid points are set to $P_{\phi} = 100$. For test mode, we fed the network with data corresponding to $P_{\text{TEST}}=100$ DoA angles different than the ones used in the training phase but keeping the values of $M$, $K$, $L$ and $T$ same as in the training. Fig.~\ref{figULABoundAnalysis}a shows the percentage of times a particular antenna index appears as part of the optimal array in the output over $J_{\text{TEST}}=T_{\text{TEST}}P_{\text{TEST}}$ trials with different performance metrics used during training. As seen here, when the CNN is trained with data created from the CRB, the classifier output arrays usually consists of the elements at the extremities. However, the network trained on BZB and WWB usually selects arrays with elements close to each other leading to low sidelobe levels. Also shown here is the random selection wherein each element is chosen with approximately $10\%$ selection rate. We provide the DoA estimation performance of the antenna subarrays selected by the network for different values of test data SNRs in Fig.~\ref{figULABoundAnalysis}b. We observe that, compared to the DL-based approach, the random thinning results in inferior DoA estimation due to small array aperture. Among various bounds, the MSE is somewhat similar at high SNR regimes with the BZB faring better than CRB at low SNRs.

\section{TL for Sparse Arrays}
\label{sec:transfer}
The CNN architecture presented in the previous section is designed for a specific array geometry and is, therefore, inapplicable to different array configurations without significant re-training with new data. This arises from the assumption that the data used for training and testing are drawn from the same or similar distribution, which is difficult to guarantee in real world. In deep learning, this problem is called \textit{domain mismatch} \cite{duan2012domain}. On the other hand, labeling sufficient training data for all possible application domains is prohibitive. It has been shown \cite{kulis2011you} that it is possible to establish a reasonable model by exploiting the labeled data drawn from another sufficiently labeled \textit{source domain} which is closer to describing similar contents of the \textit{target domain}. This \textit{domain adaptation} (DA) enables knowledge transfer across domains. Lately, TL has emerged as an effective domain adaptation technique, wherein the DL network learns domain-invariant models across source and target domains \cite{pan2010survey}, and has been applied to processing of image, 
	bio-medical, 
	radar, 
	and speech 
	signals.
	
	Apart from a domain mismatch problem, networks such as a CNN also suffer from the need of a large training database. When only limited labeled data is available, the CNN fails to optimally select the sensor subarrays. Since CNN objective functions are highly non-convex and convergence of optimization algorithms to a global optimum is not guaranteed, training with only large data could increase the probability of convergence. Alternatively, training forms such as convolutional autoencoders (CAE) \cite{vincent2010stacked} and TL are employed in data-limited applications. Some studies \cite{seyfiouglu2017deep} suggest that TL outperforms CAE especially when the sample sizes are very small.
	
	In TL-based antenna selection, we address the domain mismatch between various array geometries and lack of massive training data by developing a more efficient, deep TL-based sensor subarray selection approach. Indeed, sufficient datasets required to support the level of training CNNs need are unavailable, expensive or impossible to extract in many real-world sensor array applications. We apply TL to enable the network in selecting sensor subarrays accurately even when limited labeled data are available. In particular, we transfer the features in the training data from one array geometry to a different array configuration. For example, we use a CNN trained with a URA to select sensors in a UCA. This domain transfer is advantageous when a large off-line database is required for system identification or calibration \cite{comparisonOfMIMOArrays}. 
	
	Conventionally, DoA estimation across various geometries is performed with array transformation and array interpolation techniques \cite{ref_AI5}. However, for multiple targets and complex geometries, these techniques are difficult to come by \cite{rubsamen2008direction,liu2019doa}. TL-based approach is helpful in overcoming such limitations. 
	

	\begin{figure}[t]
		\centering
		\subfloat[][]{\includegraphics[scale=.53]{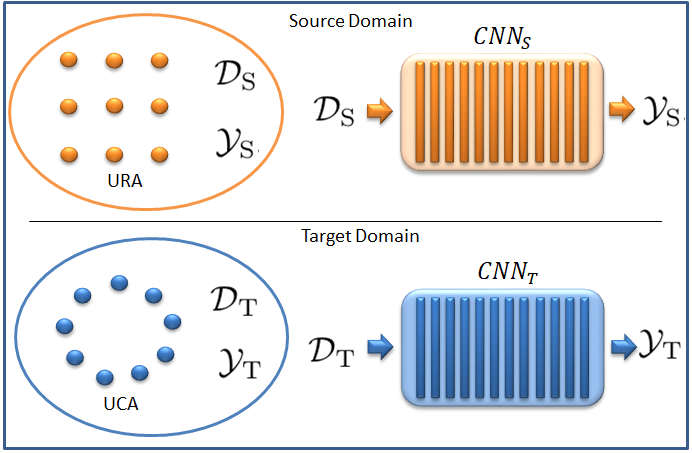}} \\
	\subfloat[][]{\includegraphics[scale=.50]{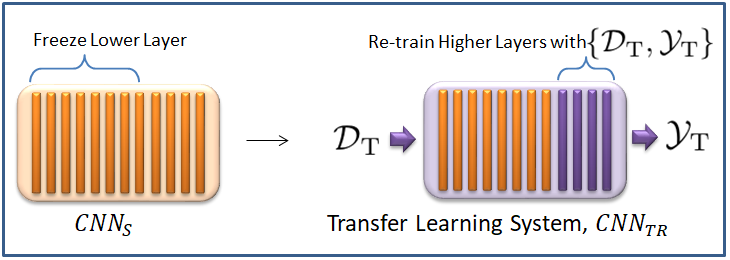}}
		\caption{The representation of the source and target domains for knowledge transfer from, for example, URA to UCA configuration. (a) Source (top) and target (bottom) domain data with corresponding learning networks $\mathrm{CNN}_\mathrm{S}$ and $\mathrm{CNN}_\mathrm{T}$, respectively. (b) In deep TL, lower layers of $\mathrm{CNN}_\mathrm{S}$ are frozen and only higher layers are re-trained with the target domain (UCA) data $\{\mathcal{D}_\mathrm{T}, \mathcal{Y}_\mathrm{T}\}$ to transfer sensor selection knowledge from source domain (URA) \cite{elbir2020TL}.}
		\label{figTransNet}
	\end{figure}

 When compared with the domain transfer in shallow TL techniques \cite{pan2010domain}, such as classification based on SVM, a deep TL approach combines DA with the power of a deep network to learn the explanatory factors of variations in data and reduce the mismatch between the marginal distributions across array geometries. In Fig.~\ref{figTransNet}, we define the source (target) data and labels as $\mathcal{D}_\mathrm{S}$ ($\mathcal{D}_\mathrm{T}$) and $\mathcal{Y}_\mathrm{S}$ ($\mathcal{Y}_\mathrm{T}$), respectively. We train the source network $\mathrm{CNN}_\mathrm{S}$, which learns the non-linear relationship between $\mathcal{D}_\mathrm{S}$ and $\mathcal{Y}_\mathrm{S}$ as
	\begin{align}
	\label{Ys}
	\mathcal{Y}_S = \Sigma_\mathrm{S} (\mathcal{D}_\mathrm{S}),
	\end{align}
	where $\Sigma_\mathrm{S}(\cdot) $ is the non-linear function that constructs the mapping between the data and labels in the source domain. In (\ref{Ys}), the label data are the positions of the best subarray sensors as $\mathcal{Y}_\mathrm{S} = \{ \mathcal{Y}_\mathrm{S}^{(1)},\mathcal{Y}_\mathrm{S}^{(2)},\dots,\mathcal{Y}_\mathrm{S}^{(\textsf{S})} \}$ where $\textsf{S} = |\mathcal{Y}_\mathrm{S} | = |\mathcal{D}_\mathrm{S} |$ is the number of elements in the source domain.
	Furthermore, $\mathcal{D}_\mathrm{S}$ is the collection of covariance matrices of the array outputs of the source array geometry, i.e.,
	\begin{align}
	\mathcal{D}_\mathrm{S} = \{ \mathcal{D}_\mathrm{S}^{(1)},\mathcal{D}_\mathrm{S}^{(2)},\dots, \mathcal{D}_\mathrm{S}^{(\textsf{S})}   \},
	\end{align}
	where $\mathcal{D}_\mathrm{S}^{(i)} = \mathbf{X}_\mathrm{S}$ which is constructed from the source domain covariance matrix
	\begin{align}
	\mathbf{R}_\mathrm{S} = \frac{1}{T} \sum_{i = 1}^{T} \mathbf{y}_\mathrm{S} (t_i) \mathbf{y}_\mathrm{S}^H(t_i),
	\end{align}
	where $\mathbf{y}_\mathrm{S}(t_i)$ denotes the array output of the source data.
	
	Similarly, the target domain data and labels are $	\mathcal{D}_\mathrm{T} = \{ \mathcal{D}_\mathrm{T}^{(1)},\mathcal{D}_\mathrm{T}^{(2)},\dots, \mathcal{D}_\mathrm{T}^{(\textsf{T})}   \}$ and $	\mathcal{Y}_\mathrm{T} = \{ \mathcal{Y}_\mathrm{T}^{(1)},\mathcal{Y}_\mathrm{T}^{(2)},\dots, \mathcal{Y}_\mathrm{T}^{(\textsf{T})}   \}$, respectively, where $\textsf{T} = | \mathcal{D}_\mathrm{T}|$ and $	\mathbf{R}_\mathrm{T} = \frac{1}{T} \sum_{i = 1}^{T} \mathbf{y}_\mathrm{T} (t_i) \mathbf{y}_\mathrm{T}^H(t_i)$. For the target network $\mathrm{CNN}_\mathrm{T}$, we have
	\begin{align}
	\mathcal{Y}_\mathrm{T} = \Sigma_\mathrm{T} (\mathcal{D}_\mathrm{T}).
	\end{align}
	The TL framework assumes that the source domain has much larger dataset than the target domain, i.e., $\textsf{S} \gg \textsf{T}$. This implies that $\mathrm{CNN}_\mathrm{S}$ will turn out to be a well-trained deep network whereas $\mathrm{CNN}_\mathrm{T}$ has poor mapping performance and does not reflect the same mapping profile as $\mathrm{CNN}_\mathrm{S}$. To improve the performance of $\mathrm{CNN}_\mathrm{T}$, the key idea is to use the sensor selection ability of the pre-trained network $\mathrm{CNN}_\mathrm{S}$ even if it is trained with different array data~\cite{freeezingLayers3}. \textcolor{black}{This is achieved by re-training $\mathrm{CNN}_\mathrm{S}$ with the target domain data $\{\mathcal{D}_\mathrm{T},\mathcal{Y}_\mathrm{T}\}$ while freezing the lower layers (i.e., convolutional layers) of $\mathrm{CNN}_\mathrm{S}$\footnote{\color{black}We do not freeze the layers $\{3, 5, 7\}$, because they are $\mathrm{ReLU}$ layers with no weight to freeze.}. The new deep transfer network is $\mathrm{CNN}_\mathrm{TR}$ (Fig.~\ref{figTransNet}b).} The lower layers are kept intact or \textit{frozen} because they are generally domain invariant\footnote{``Domain invariance" implies that when new labels are added to the network, the lower layers remain unaffected even though the problem has changed.} and hence, harbor the bulk of sensor selection knowledge. The higher layers, however, are largely domain variant such that when new labels are added to the problem (i.e., $\mathcal{Y}_\mathrm{S}$ is replaced with $\mathcal{Y}_\mathrm{T}$), they require re-training. {\color{black}This approach  accelerates the computation of  the gradient in the backpropagation stage. Furthermore, it allows us to enlarge the feature space of the deep network without causing large error on the already-learned features~\cite{freeezingLayers3}.}
	\\\\
	\noindent\textbf{Knowledge Transfer Across Different Array Geometries} 
	Once $\mathrm{CNN}_\mathrm{S}$ (i.e., $\Sigma_\mathrm{S}(\cdot)$) is trained with the source domain data, we freeze the weights in the $\{2, 4, 6, 8\}$-th layers (i.e., the convolutional layers) to preserve the sensor selection ability of the deep network before transferring it to the target domain. We construct the TL network such that
	\begin{align}
	\Sigma_\mathrm{TR} (\mathcal{D}_\mathrm{T}) =  f^{(15)} \big(f^{(14)} ( \dots \tilde{f}^{(2)}(f^{(1)}( \mathcal{D}_\mathrm{T} )   )   )   \big)  = \mathcal{Y}_\mathrm{T},\label{TLmapping}
	\end{align}
	where the frozen layers are $\tilde{f}^{(i)}_{i \in \{2, 4, 6, 8\}}$. Algorithm~\ref{alg:TL} lists these steps of the TL approach.
	\begin{algorithm}[t]
		\begin{algorithmic}[1]
			\caption{Transfer learning for sensor selection.}
			\Statex {\textbf{Input:} $\{\mathcal{D}_\mathrm{S}, \mathcal{Y}_\mathrm{S}\}$,  $\{\mathcal{D}_\mathrm{T}, \mathcal{Y}_\mathrm{T}\}$ }
			\label{alg:TL}
			\Statex {\textbf{Output:} $\mathrm{CNN}_\mathrm{TR}$.}
			\State Train $\mathrm{CNN}_\mathrm{S}$ with $\{\mathcal{D}_\mathrm{S}, \mathcal{Y}_\mathrm{S}\}$.
			\State Construct TL network $\mathrm{CNN}_\mathrm{TR}$ whose convolutional layers are designated the same as of $\mathrm{CNN}_\mathrm{S}$, i.e., ${f_\mathrm{TR}}^{(i)}_{i \in \{2, 4, 6, 8\}} = {f_\mathrm{S}}^{(i)}_{i \in \{2, 4, 6, 8\}}$.
			\State Train the remaining layers of the TL network with $\{\mathcal{D}_\mathrm{T}, \mathcal{Y}_\mathrm{T}\}$. Then, use $\mathrm{CNN}_\mathrm{TR}$ for sensor selection for target domain data.
		\end{algorithmic}
	\end{algorithm}
\\\\
\noindent\textbf{Deep Network Realization and Training}
	 For training, we used stochastic gradient descent algorithm with momentum $0.9$ and updated the network parameters at learning rate $0.01$ and mini-batch size of $512$. The loss function was the cross-entropy cost\par\noindent\small
	\begin{align}
	\label{costFunction}
	\mathrm{C}_\mathrm{E} = -\frac{1}{\bar{\textsf{T}}} \sum_{t = 1}^{\bar{\textsf{T}}} \sum_{c = 1}^{\bar{C}}  \bigg[\chi_c^{(t)} \ln \eta_c^{(t)} + (1 - \chi_c^{(t)}) \ln (1 - \eta_c^{(t)})  \bigg],
	\end{align}\normalsize
	where $\bar{\textsf{T}}$ is the length of the dataset and $\{\eta_c^{(t)}, \chi_c^{(t)}  \}_{t = 1, c=1}^{\bar{\textsf{T}}, \bar{C}}$ is the input-output pair for the classification layer. {\color{black}It is worth noting that the cost function in (\ref{costFunction}) can be defined in terms of RMSE of DoA estimation procedure. However, this makes the training process problem-dependent.} During training, the training data is shuffled for each epoch until training is terminated. Further, $80\%$ and $20\%$ of all generated data are chosen for training and validation datasets, respectively. The training rate is reduced by a factor of $0.9$ after each $10$ epochs. The training stops when the validation accuracy does not improve for three consecutive epochs.

 To train $\mathrm{CNN}_\mathrm{S}$, we collected array data for $P_\mathrm{S}=100$ equally spaced direction in the sector $\tilde{\Theta} = [0^\circ, 359^\circ]$ azimuth plane and $L_\mathrm{S}=100$ noisy data realizations with $T=100$ data snapshots. During training, we set $\sigma_s^2=1$ and use different SNR levels, namely, $\mathrm{SNR}_\mathrm{TRAIN} \in \{15, 20, 25\}$ dB. Hence, the total training data length is $3L_\mathrm{S}P_\mathrm{S}=30000$. Once $\mathrm{CNN}_\mathrm{S}$ is trained, the $\mathrm{CNN}_\mathrm{TR}$ is constructed by following the steps in Algorithm~\ref{alg:TL}. For the above-mentioned settings with $M=16$ and $K=6$, the training time for $\mathrm{CNN}_\mathrm{S}$, $\mathrm{CNN}_\mathrm{T}$ are approximately $40$ and $5$ minutes respectively, whereas the TL network  $\mathrm{CNN}_\mathrm{TR}$ needs only $5$ seconds to be trained.

	\begin{figure}[t]
		\centering
		\includegraphics[width=.35\textheight]{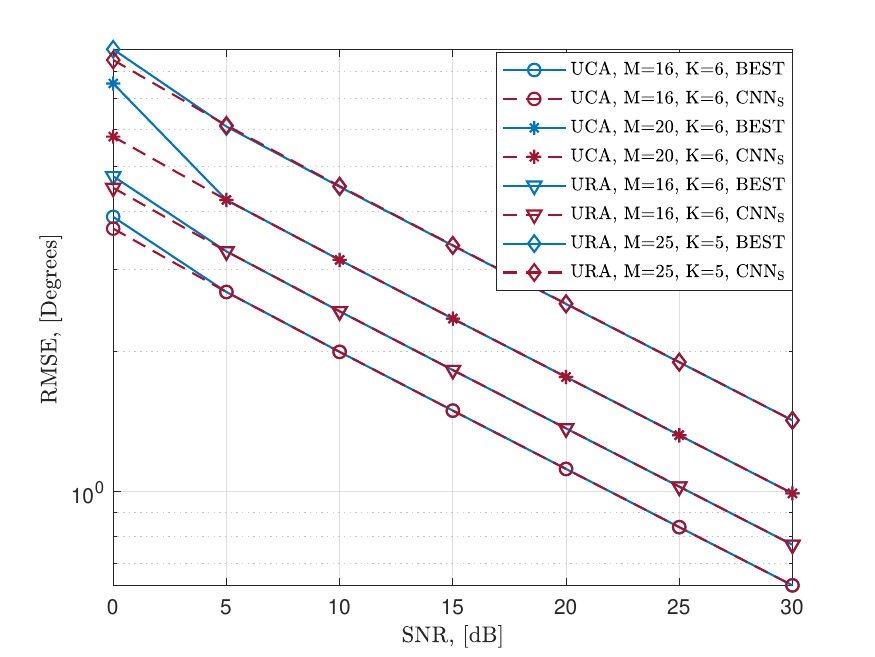} 
		\caption{DoA estimation performance for $\mathrm{CNN}_\mathrm{S}$ for  different array geometries \cite{elbir2020TL}.}
		\label{figSourceDOAAll}
	\end{figure}
	
 \subsection{Performance in Source Domain}
We first present the performance of the  CNN approach for the source domain case where different array geometries such as URA and UCA are considered with different array settings. In particular, we consider sensor arrays with half wavelength sensor spacing for both UCA and URA. When $\mathrm{CNN}_\mathrm{S}$s are trained for different arrays, we obtained above $90\%$ validation accuracy for the training data in all cases. In the prediction stage, the DoA angles are generated uniformly at random in the interval $\tilde{\Theta}$ so that the DoA angles in the training and prediction are selected from the same distribution. After feeding $\mathrm{CNN}_\mathrm{S}$ with these input data, the selected subarrays are obtained from the output for each scenario. Then, the sensor outputs of corresponding subarrays are employed for DoA estimation using MUSIC (MUltiple SIgnal Classification) algorithm~\cite{music}. During the simulations in the prediction state, the network is tested for different SNR levels for $J_T=100$ Monte Carlo trials. Figure~\ref{figSourceDOAAll} shows the RMSE in DoA estimation, i.e.,
	\begin{align}
	\mathrm{RMSE} = \bigg(\frac{1}{J_T  } \sum_{j=1}^{J_T} (\hat{\phi}^{(j)} - \phi)^2 \bigg)^{\frac{1}{2}},
	\end{align}
	where $\hat{\phi}^{(j)}$ and $\phi$ denote the estimated and true DoA angles, respectively. We compare the DoA estimation performance of $\mathrm{CNN}_\mathrm{S}$ with the best subarray that provides the lowest CRB. Figure~\ref{figSourceDOAAll} demonstrates that $\mathrm{CNN}_\mathrm{S}$ asymptotically follows the best subarray performance. 

 \subsection{Performance for Transfer Learning}

	\begin{figure}[t]
		\centering
		\subfloat[][]{\includegraphics[width=.46\textwidth]{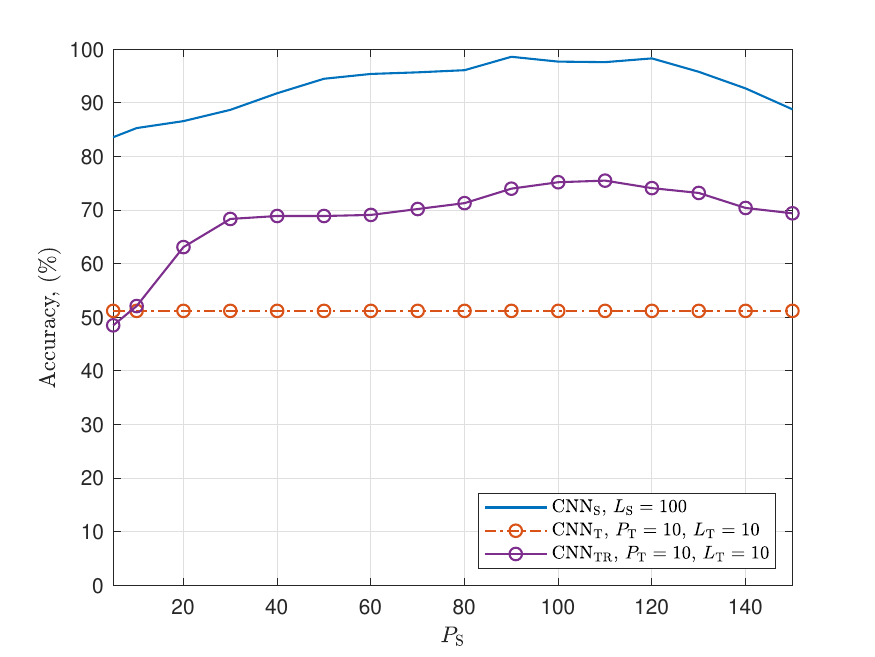}}
		\subfloat[][]{\includegraphics[width=.46\textwidth]{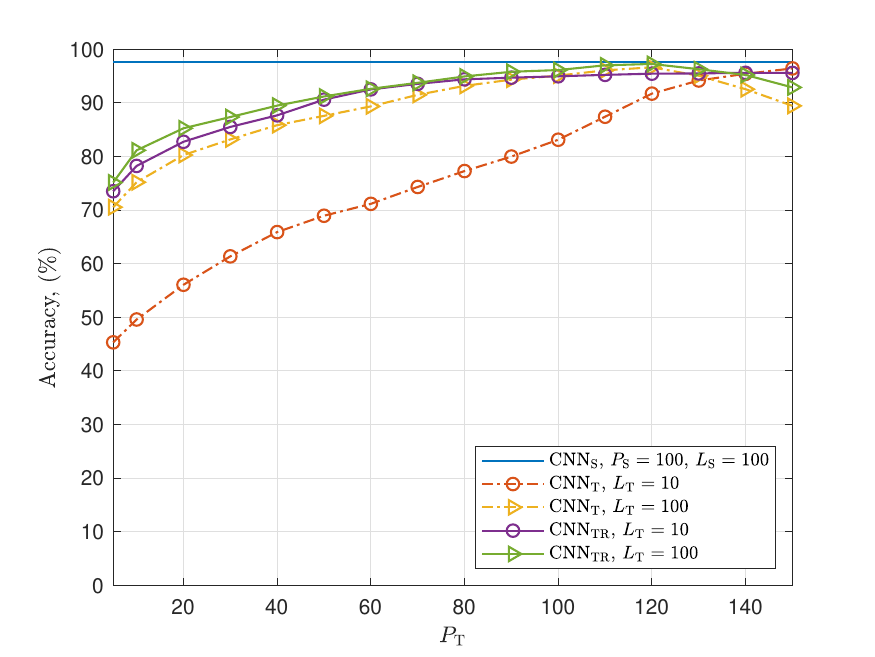}}
		\caption{\textcolor{black}{Performance of $\mathrm{CNN}_\mathrm{S}$, $\mathrm{CNN}_\mathrm{T}$ and $\mathrm{CNN}_\mathrm{TR}$  versus the number of DoA angles. Sensor selection accuracy is given with respect to (a) $P_\mathrm{S}$  when $L_\mathrm{S}=100$, $L_\mathrm{T}=10$, and (b) $P_\mathrm{T}$ when $P_\mathrm{S} = 100$, $L_\mathrm{S}=100$. $\mathrm{SNR}_\mathrm{{TRAIN}}=15$ dB \cite{elbir2020TL}.}}
		\label{figTLTest}
	\end{figure}

	In order to evaluate the TL performance, we trained $\mathrm{CNN}_\mathrm{S}$ with different sizes of datasets and then constructed $\mathrm{CNN}_\mathrm{TR}$ from $\mathrm{CNN}_\mathrm{S}$ for sensor selection. We considered URA and UCA geometries with $M=16$, $K=6$ for source and target domains, respectively. Fig.~\ref{figTLTest} shows the sensor selection accuracy
		\begin{align}
		\mathrm{Accuracy}(\%) = \frac{\textsf{U}}{\textsf{V}}\times 100,
		\end{align} 
		where $\textsf{V}$ is the total number of input datasets in which the model identified the best subarrays correctly $\textsf{U}$ times. In Fig.~\ref{figTLTest}a, the target domain $\mathcal{D}_\mathrm{T}$ are generated for $P_\mathrm{T}=10$ grid points in $\widetilde{\Theta}$ and $L_\mathrm{T}=10$ and we varied $P_\mathrm{S}$ from $5$ to $150$ for $\mathcal{D}_\mathrm{S}$ with $L_\mathrm{S}=100$. For all three networks,  The performance of $\mathrm{CNN}_\mathrm{T}$ is fixed because $\mathcal{D}_\mathrm{T}$ does not change during the simulations. When $P_\mathrm{S}$ is very small (i.e., $<10$), $\mathrm{CNN}_\mathrm{TR}$ performs even worse than $\mathrm{CNN}_\mathrm{T}$. However, as $P_\mathrm{S}$ increases, $\mathrm{CNN}_\mathrm{TR}$ and $\mathrm{CNN}_\mathrm{S}$ exhibit higher selection accuracy. For large source datasets, e.g. $P_\mathrm{S}\in [80, 120]$, $\mathrm{CNN}_\mathrm{TR}$ outperforms $\mathrm{CNN}_\mathrm{T}$ by a large margin because of the learned and transferred features from $\mathrm{CNN}_\mathrm{S}$. The increase in $P_\mathrm{S}$ does not necessarily improve the sensor selection performance because when the training data are densely sampled (i.e., $P_\mathrm{S}$ is high) the deep network cannot distinguish the input data of different directions and produce inaccurate classification output. These results suggest that $\mathrm{CNN}_\mathrm{S}$ needs to be trained with at least $\textsf{S} = P_\mathrm{S}L_\mathrm{S} = 40\cdot 100=4000$ to provide satisfactory accuracy (e.g., above $90\%$). As a result, $P_\mathrm{S} = 100$ is a reasonable choice for TL, wherein the target dataset $1000$ times smaller, i.e., $\frac{\textsf{S}}{\textsf{T}} = \frac{L_\mathrm{S}P_\mathrm{S}}{L_\mathrm{T}P_\mathrm{T}} = 1000$. In Fig.~\ref{figTLTest}b, we repeat the same analysis for $\mathrm{CNN}_\mathrm{T}$ where we assume that $\mathrm{CNN}_\mathrm{S}$ is well-trained with $P_\mathrm{S}=100$ and $L_\mathrm{S}=100$. Then, we sweep $P_\mathrm{T}$ similarly for both $L_\mathrm{T}=10$ and $L_\mathrm{T}=100$. We can see that when $L_\mathrm{T} = 100$, $\mathrm{CNN}_\mathrm{T}$ quickly reaches maximum similar to $\mathrm{CNN}_\mathrm{S}$ as illustrated in Fig.~\ref{figTLTest}a. In this case, the improvement gained by TL is incremental because $\mathrm{CNN}_\mathrm{T}$ is already well-trained. However, if small dataset is used, i.e., $L_\mathrm{T}=10$, then it requires larger $P_\mathrm{T}$ to reach high accuracy. Expectedly, this analysis shows that TL provides reasonable improvement if the target dataset is relatively small, i.e., $\textsf{T} = P_\mathrm{T}L_\mathrm{T}\leq 1000$ ($\textsf{S} = 10000$). In other words, when $\textsf{T}$ is high there is no need to use TL. Therefore, in the following experiments, we select $P_\mathrm{T}=L_\mathrm{T}=10$ and employ TL to improve the performance. 

	\begin{table}[t]
		\caption{ Training Validation Accuracy (\%) For Different TL Scenarios		\label{tableValAcc}}
		{\begin{tabular}{|c|c|c|}
			\hline
			\multirow{2}{*}{TL Scenario (Source $\rightarrow$ Target)}  & \multicolumn{2}{|c|}{Validation Accuracy (\%)} \\
			&         $\mathrm{CNN}_\mathrm{T}$    &  $\mathrm{CNN}_\mathrm{TR}$              \\
			\hline
			UCA $\rightarrow $ URA, $M=16$, $K=6$  &	54.9 & 	70.1	 \\
			\hline
			URA $\rightarrow $ UCA, $M=16$, $K=6$  &	42.3 & 	79.8	 \\
			\hline
			UCA $\rightarrow $ $\overline{\mathrm{UCA}}$, $M=20$, $K=6$  &	63.1 & 	98.8	 \\
			\hline
			URA $\rightarrow $ $\overline{\mathrm{URA}}$, $M=25$, $K=5$  &	55.2 & 	77.4	 \\
			\hline 
		\end{tabular}}{}
	\end{table}

Table~\ref{tableValAcc} lists the validation accuracy of $\mathrm{CNN}_\mathrm{T}$ and $\mathrm{CNN}_\mathrm{TR}$ for different TL scenarios. We consider TL between UCA and URA as well as the perturbed array geometries denoted by $\overline{\mathrm{UCA}}$ and $\overline{\mathrm{URA}}$. In a perturbed array geometry, the $m$-th sensor position is selected uniformly at random as $\{\tilde{x}_m, \tilde{y}_m,\tilde{z}_m\} \sim \mathcal{N}(\{x_m, y_m, z_m\},(\lambda/4)^2)$ for each instance of the training data. It is evident that the sensor selection accuracy of $\mathrm{CNN}_\mathrm{TR}$ is approximately $20\%$ higher than $\mathrm{CNN}_\mathrm{T}$.

	\begin{figure}[t]
		\centering
		\includegraphics[width=.5\textwidth]{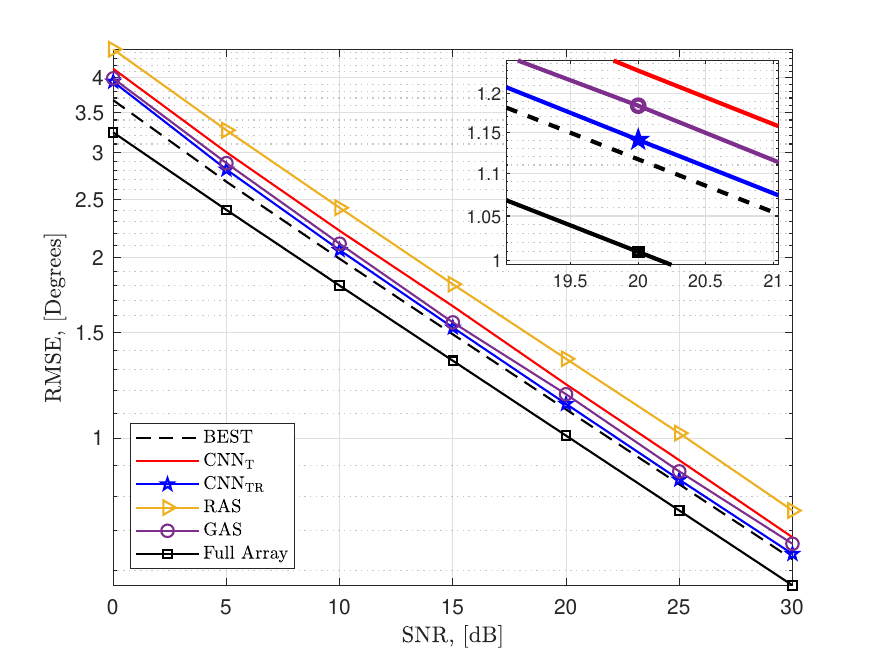} 
		\caption{DoA estimation performance when source domain has URA; target domain has UCA geometry when $M=16$, $K=6$ \cite{elbir2020TL}.}
		\label{figTL_URA_UCA}
	\end{figure}
\begin{figure}[t]
		\centering
		\includegraphics[width=.5\textwidth]{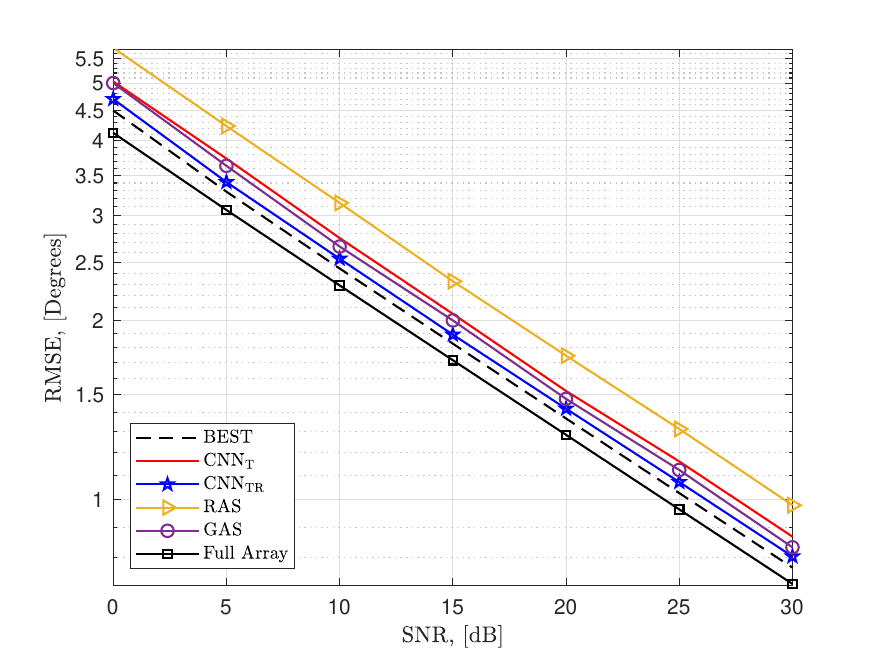} 
		\caption{DoA estimation performance when source domain has UCA; target domain has URA geometry when $M=16$, $K=6$ \cite{elbir2020TL}.}
		\label{figTL_UCA_URA}
	\end{figure}

	We further assessed the DoA estimation performance of the selected subarrays 
	for target domain data. For $M=16$, $K=6$, Figs.~\ref{figTL_URA_UCA} and \ref{figTL_UCA_URA} depict the performance for URA$\rightarrow$UCA and UCA$\rightarrow$URA scenarios, respectively. We compared the sensor selection performance of $\mathrm{CNN}_\mathrm{T}$ and $\mathrm{CNN}_\mathrm{TR}$ {\color{black}with  greedy-based antenna selection (GAS)~\cite{antennaSelectionKnapsack}, random selection (RAS) as well as the fully array performance. As expected, we see that the full array has the lowest SNR due to large array aperture. We observe that $\mathrm{CNN}_\mathrm{TR}$ closely follows the performance of the best subarray. The $\mathrm{CNN}_\mathrm{TR}$ exhibits approximately $4\%$, $8.5\%$ and $23\%$ lower RMSE as compared to GAS, $\mathrm{CNN}_\mathrm{T}$ and RAS, respectively. It is worth noting that RAS has no rule on selecting the antennas while GAS is a greedy-based suboptimum method seeking the best subarray based on the CRB information~\cite{antennaSelectionKnapsack}.} These results establish the effectiveness of TL for DoA estimation with sensor selection. The superior performance of $\mathrm{CNN}_\mathrm{TR}$ is because of the learned and transferred features from source domain data via $\mathrm{CNN}_\mathrm{S}$.

\section{Large Planar Sparse Array Design with SA-Assisted DL}
\label{sec:2d}
The SA-assisted DL-based approach was originally presented in \cite{ieice2021,isap2021} to optimize sensor positions while minimizing the mutual coupling between the sensors. It was extended to 2-D sparse array design in \cite{access2021}. Here, given an initial array as $\mathcal{Z}_{init}$, the optimized array $\mathcal{Z}_{SA}$ is obtained by minimizing
\begin{eqnarray}
\mathcal{Z}_{SA}
&=&
\argmin_{\mathcal{Z}_{init}} {k_o}=\sum_{i=1}^M \sum_{j=i+1}^M \frac{1}{||m_i-m_j||_2}, \\[10pt]
\label{Cost_1}
&& {\text{such that}} ~~~~~~ {||m_i-m_j||_2} \leqslant B, \nonumber
\label{cost}
\end{eqnarray}
where $m_i,m_j \in \mathcal{Z}_{init}$, $M$ is the number of sensors, $|| \cdot||_2$ is the
$l_2$-norm of a vector and $B$ is the mutual coupling coefficient upper-bound.
The SA algorithm is used as an initialization step for the deep learning model, mainly to generate sparse subarrays with large physical apertures and well-distributed sensors instead of an enumeration approach.

\begin{figure}[t]
\centering
\includegraphics[width=50mm]{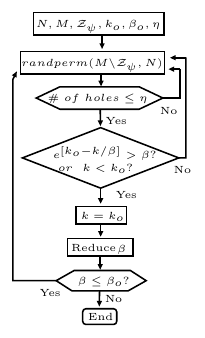}
\caption{Improved SA algorithm for random $2\text{D}$ sparse subarray generation and optimization.}
\label{algo}
\end{figure}

Figure \ref{algo} summarizes the step-by-step procedure of the flow diagram. 
Note that this SA-based approach can be extended to any planar array configuration, and the same applies to the whole SA-based initialization method \cite{access2021}--\cite{isap2021}.
The best subarrays are selected from ${\mathcal{H}}_{sa}$ with the lowest CRB values as labels. 

The whole algorithm is summarized in the Algorithm \ref{B}. 
The inputs are as follows: the total number of given antennas $M$, the number of antennas to be selected $K$, the number of snapshots $T$, the number of different DOA angles $D$, the number of signals and noise realizations $P$ and the SNR. 
Moreover, the elements of $\mathcal V$ are chosen from $\mathcal{H}_{sa}$ which is calculated using the  SA-based optimization method as shown in Fig. \ref{algo} rather than the enumeration of the entire combinations like in $\mathcal{H}$. 
The SNR used for calculation of the covariance matrices is denoted as $\text{SNR}_{\text{TRAIN}}$.

\begin{algorithm}[t!]
 \caption{Training Dataset Generation Method}
 \begin{algorithmic}[1]
\Statex \textbf{Input:} $N$, $M$, $T$, $K$, $P$ and $\text{SNR}_{\text{TRAIN}}$
\Statex \textbf{Output:}  
 {Training data $\bm{\mathcal{D}}_{\text{TRAIN}}$}
  \State Generate $\mathcal{H}_{sa}$ as shown in Fig. \ref{algo}.
 \State  Sample $D$ DOA angles $\{\Theta_{d}\}_{d=1}^{D}$.
  \State Compute $P$ different realizations of subarray output, $\{\bm{X}_{d}^{i}\}_{i=1}^{P}$ for $d=1, \ldots,d$ \\ 
~~~~~~~~~~~~~~$ \bm{X}_{d}^{i}= [ \textbf{x}_{d}^{i}(1), \textbf{x}_{d}^{i}(2), \ldots, \textbf{x}_{d}^{i}(T)] $,\\
where $\textbf{x}_{d}^{i}(t)=\bm{a}(d) \bm{s}^{(i)}(t) + \bm{n}^{(i)}(t)$, \\
$\bm{s}^{(i)}(t)  \thicksim \mathcal{CN}(0, \sigma_{s}^{2} \bm{I})$ and $\bm{n}^{(i)}(t) \thicksim \mathcal{CN}(0, \sigma_{n}^{2})$
 \State Calculate sample covariance matrix $\bm{\hat R}$ and $N \times N$ covariance matrices $R^{(i,d)}_{h_s}$ for ${h_s}=1, 2, \ldots,{\bar{\mathcal H}_{sa}}$.
 \State Compute $\mathcal{C} \big(\Theta, \mathcal{Z}_{h_s} \big)$ for all ${h}_s \in {\mathcal H}_{sa}$ and select subarrays.
 \State  Create input-output data pairs as $(\bm{\hat R}^{(i,d)}, u_d^{(i)})$ for $d=1, 2, \ldots, D$ and for $i=1, 2, \ldots,P$. 
  \State Connect the input-output pairs to form the training dataset as
 \\	~~$\bm{\mathcal{D}}_{\text{TRAIN}}=\Big [ (\bm{\hat R}^{(1,1)}, u_1^{(1)}), (\bm{\hat R}^{(2,1)}, u_1^{(2)}), \ldots$\\
~~~~~~~~~~~$, (\bm{\hat R}^{(P,1)}, u_1^{(P)}), (\bm{\hat R}^{(1,2)}, u_2^{(1)}), \ldots, (R^{(P,D)}, u^{P}_{D}) \Big]$
 \\ where the size of the training dataset is $\mathcal{R} = PD$. 
 \end{algorithmic} 
 \label{B}
 \end{algorithm}

The performance of the SA-assisted DL-based antenna selection approach is evaluated through simulation. 
The problem of selecting $K=16$ sensors out of a $N = 42 = 6 \times 7$ sensor URA is considered. 
The details of training and testing specifications can be found in \cite{access2021}.

\begin{figure}[t]
\centering
\includegraphics[width=0.95\linewidth]{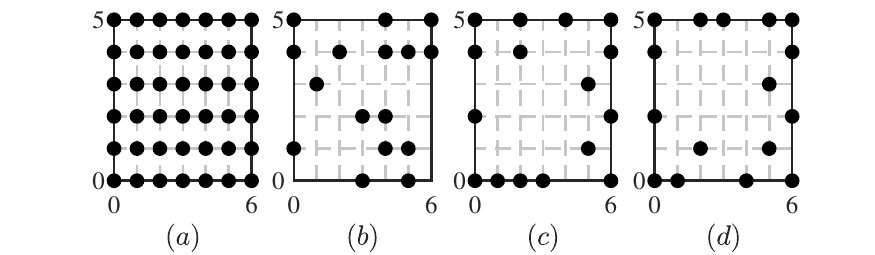} 
\caption{The array configurations of (a) parent $42-$sensor URA, (b) the conventional DL-based array ($16-$sensors),  (c) the conventional SA-based array ($16-$sensors) and (d) the  DL-based array  ($16-$sensors). Note that the dots denote physical sensors.}.
\label{arrays}
\end{figure}

The estimation accuracy of the SA-assisted DL-based method is evaluated in comparison to the conventional approach using the CNN models trained.
Figure \ref{arrays} shows the array configuration of predicted 2-D sparse arrays. Particularly, Fig. \ref{arrays} (a) illustrates the parent $42-$element URA whereas Fig. \ref{arrays} (b)--(d) show the realized $16$-element DL-based $2\text{D}$ sparse array using the conventional and DL-based method respectively. 

Here, the performance of the realized 2-D sparse arrays is evaluated. In particular, we explore the behavior of the RMSE as a function of SNR and the number of snapshots.  The RMSE is defined as
\begin{eqnarray}
\text{RMSE}
&=&
\sqrt{\frac{1}{\mathcal{T}D}\sum_{i=1}^{\mathcal{T}}\sum_{d=1}^{D}\big[(\tilde{\phi}^{i}_d-\phi_d)+(\tilde{\theta}^{i}_d-\theta_d)\big]^2},
\label{rmse2}
\end{eqnarray}
where $\tilde{\phi}^{i}_d, \tilde{\theta}^{i}_d$ and $\phi_d, \theta_d$ denote the estimated and true $d-$th DOAs in the $i-$th trial, respectively.

Table \ref{tab:spec3} lists the parameters used to compute RMSE with respect to SNR (Example \#1) and the number of snapshots (Example \#2), respectively. Moreover, the 2-D-ESPRIT algorithm is used to estimate the sources. However, if the DCA of the realized 2-D arrays has holes in the coarray, the resulting virtual 2-D array becomes irregular. Therefore, the spatial-smoothing DOA estimation method such as 2-D-ESPRIT cannot be applied.  This is the case as the 2-D-ESPRIT algorithm requires a URA array structure for spatial smoothing pre-processes. As a result, a nuclear norm minimization (NNM) approach \cite{iscas2019} is applied to fill the holes to restore a standard 2-D configuration.

\begin{figure}[t]
\centering
\begin{minipage}{0.48\linewidth}
\centering
\includegraphics[width=0.95\linewidth]{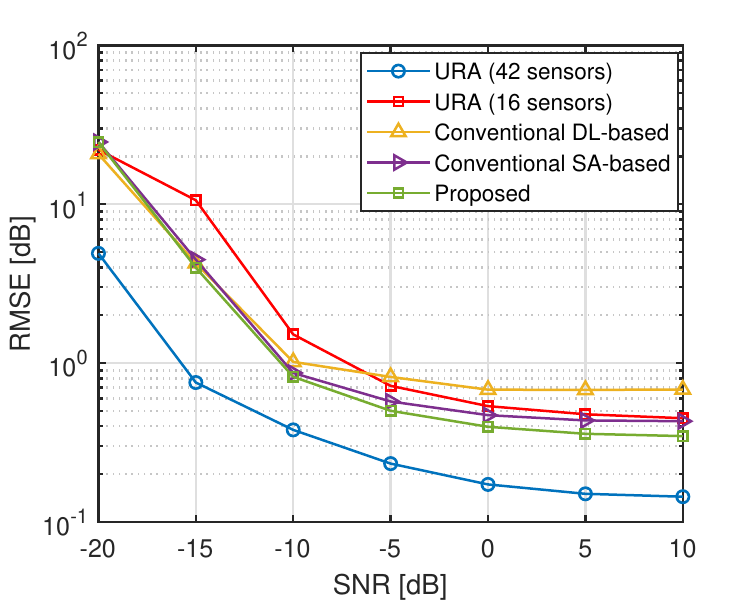} 
\\
\small{(a)}
\end{minipage}
~
\begin{minipage}{0.48\linewidth}
\centering
\includegraphics[width=0.95\linewidth]{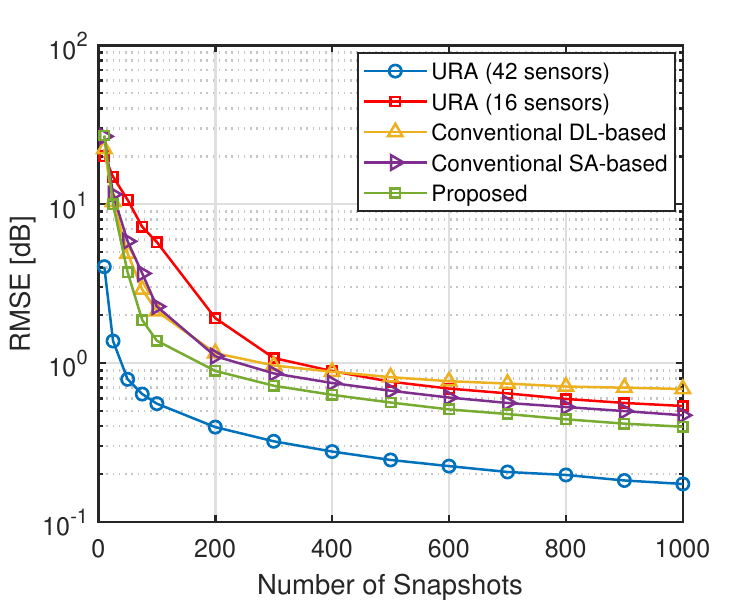}
\\
\small{(b) }
\end{minipage}
\caption{\textcolor{black}{RMSE performance versus (a) SNR and (b) number of snapshots for the realized DL-based sparse array compared to URA, conventional DL-based array and SA-based array.}} 
\label{application}
\end{figure}

Figure \ref{application} shows the DOA estimation performance of the realized DL-based $2\text{D}$ sparse array compared to the parent URA and other $2\text{D}$ sparse arrays. It can be observed in Fig. \ref{application} (a) that the URA with $42$ sensors has better performance overall due to the large physical aperture. In contrast, the performance of the DL-based $2\text{D}$ sparse array realized using the DL-based method has better than a URA with $16$ sensors and slightly lower than that of the parent URA. Besides, the   $2\text{D}$ sparse array performs better than the conventional SA-based sparse array. Moreover, the conventional DL-based performed poorly as compared to both the DL-based array and $16-$sensor URA. 

\begin{table}[ht]
\caption{2-D DOA Estimation Simulation Parameters
\label{tab:spec3} }
{\begin{tabular}{|c|c|c|}
\hline
\multirow{2}{*}{Parameters} & \multicolumn{2}{c|}{Simulation Examples} \\ \cline{2-3}
                   & Example \#1 & Example \#2 \\ \hline
\# of sources $D$       & \multicolumn{2}{c|}{9}   \\ \hline
\# of trials, $\mathcal{T}$     & \multicolumn{2}{c|}{1000}   \\ \hline
\# of snapshots & $ 500$ & $10$ to $1000$\\ \hline
SNR [dB] & $-20$ to $10$ & $0$ \\ \hline
DOAs $(\theta, \phi)$ [deg.] & \multicolumn{2}{c|}{$(10, 255), (45, 300), (20, 345), (50, 210),$} \\ 
& \multicolumn{2}{c|}{ $(70, 0), (0, 30), (60, 165), (5, 120), (30, 79) $} \\ 
\hline
Estimator                 & \multicolumn{2}{c|}{2-D-ESPRIT}   \\ \hline
\end{tabular}}{}
\end{table}

In  Fig. \ref{application} (b), we can observe a similar trend where the performance of the DL-based array is bounded by the parent URA and the $16-$sensor URA as the number of snapshots increases. 
The parent URA has the lowest RMSE values, whereas the $16-$sensor URA has higher RMSE values than the DL-based array. 
Also, the conventional DL-based array performs better than the $16-$sensor URA in lower numbers of snapshots. 
However, the RMSE values degrade as the number of snapshots increases.  
These examples demonstrate that the DL method can thin 2-D arrays cognitively to a manageable size without considerable loss of DOA estimation resolution. 


    \begin{figure}[t]
    \centering
    \includegraphics[width=1.0\textwidth]{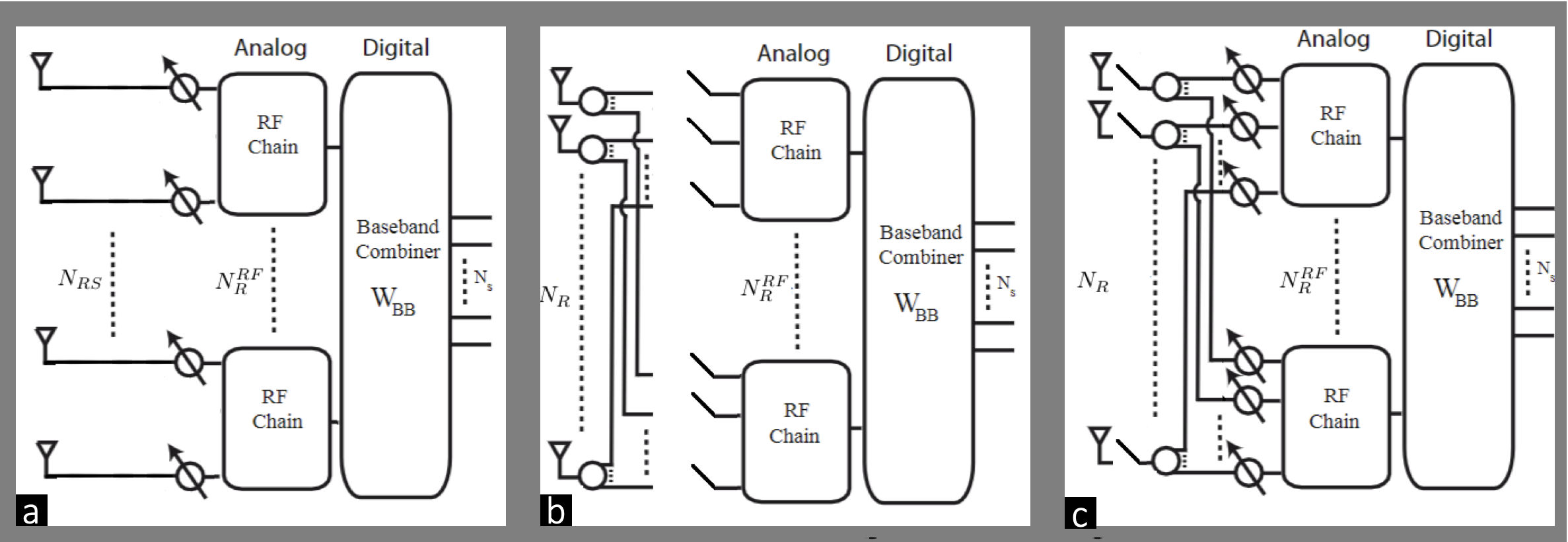} 
    \caption{\scriptsize{Receiver architectures with antenna selection for single user mmWave MIMO systems. (a) Scheme 1: Fixed subarray with  fully-connected phase shifters. (b) Scheme 2: Switching network without optimized antenna selection and no phase shifters. (c) Scheme 3:	Switching network with optimized antenna selection and phase shifters \cite{elbirQuantizedCNN2019}. }
    \vspace{-12pt}}
    \label{figAntennaSelectionSchemes}
    \end{figure}

\section{DL-based Sparse Array Design for Hybrid Beamforming}
\label{sec:beamforming}
 DL may also be used to simultaneously select antenna elements and design hybrid beamformers for massive MIMO wireless communications~\cite{elbirQuantizedCNN2019}. Here, we employ a CNN for each task. As before, the element selection problem is cast as a classification~\cite{elbirAPS2019}. We then include the hybrid beamformer design in this DL framework by exploiting the structure of analog beamformers  which are obtained by minimizing the cost between hybrid and unconstrained beamformers. The optimization problem is cast jointly with the antenna selection problem and it is solved  by MATLAB-based Manopt algorithm \cite{manopt} via manifold optimization (MO) \cite{hybridBFAltMin}.

 Among the prevalent architectures for sparse array beamformers, a popular scheme is to employ a predetermined subarray with $N_{RS}$ antennas selected from a full array of $N_R$ elements (Fig.~\ref{figAntennaSelectionSchemes}a). Each subarray feeds into a fully-connected phase shifter network of size $N_R^{RF}$ with a single RF chain. This has the complexity of phase shifters but the antenna selection process is not optimized. Another common receiver architecture feeds the antennas directly to the RF chains thereby eliminating the phase shifters completely. Here, each RF chain is connected to the $N_{R}$ antennas of which $N_{RS}$ elements are selected using switches (Fig.~\ref{figAntennaSelectionSchemes}b).  In this case, the entries of the combiner matrix are either 1 or 0 to indicate the selected or unselected antennas, respectively. This is the simplest structure with no phase shifters. However, the antenna selection is not optimized and the elements are determined by simply choosing the largest absolute values in each column of the channel matrix. Finally, Fig.~\ref{figAntennaSelectionSchemes}c shows a receiver that employs a switching network with phase shifters. In this system, a subarray with $N_{RS}$ antennas is selected from a full array comprising $N_{R}$ antennas. The subarray is connected to a phase shifter network of size $N_R^{RF}$ which may apply an optimization procedure for antenna selection to achieve greater efficiency.
 
\begin{figure*}[t]
		\centering
		{\includegraphics[width=1.0\textwidth]{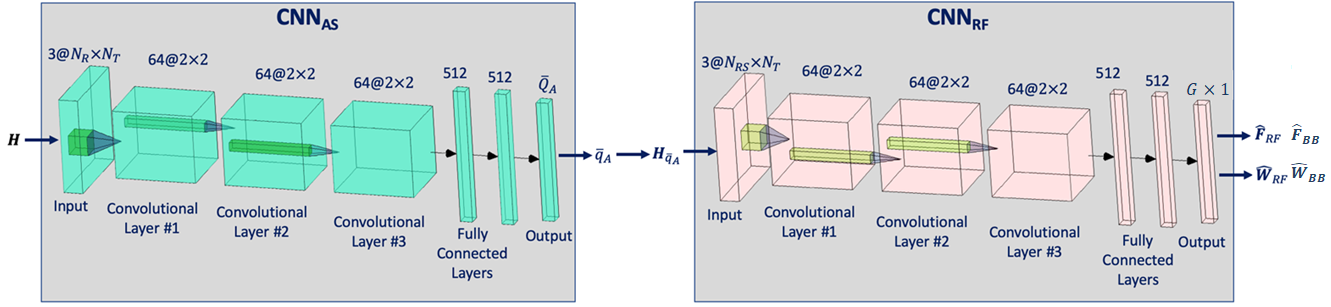}}
		\caption{The  CNN architecture for antenna selection and RF beamformer design \cite{elbirQuantizedCNN2019}. }
		\label{figNetwork2}
	\end{figure*}

	In this formulation, a CNN accepts channel matrix as input and provides the subarray that maximizes the spectral efficiency. Once the antenna selection  (CNN$_\mathrm{AS}$) is finalized (see Fig.~\ref{figNetwork2}), the corresponding partial channel matrix is fed to a second CNN (CNN$_\mathrm{RF}$) which then chooses the best RF beamformer and constructs the corresponding baseband beamformer. To train both CNN models, different realizations of the channel matrix are used and the input data are labeled by the selected subarray/RF chains with the highest spectral efficiency. Even though DL network structures require channel matrix as an input, precise knowledge of this matrix is  only required in the training stage to obtain the labels of the network. In the prediction stage, where the RF beamformers are estimated, precise channel knowledge is not necessary. Both CNNs are trained with channel matrices generated for different user location, channel gains and number of user clusters. Furthermore, each realization of channel matrix in the training data is corrupted by synthetic noise so that the performance of the learning network does not deteriorate with noisy test inputs. 

	\begin{figure*}[ht]
		\centering
		\subfloat[][]{\includegraphics[width=.45\textwidth]{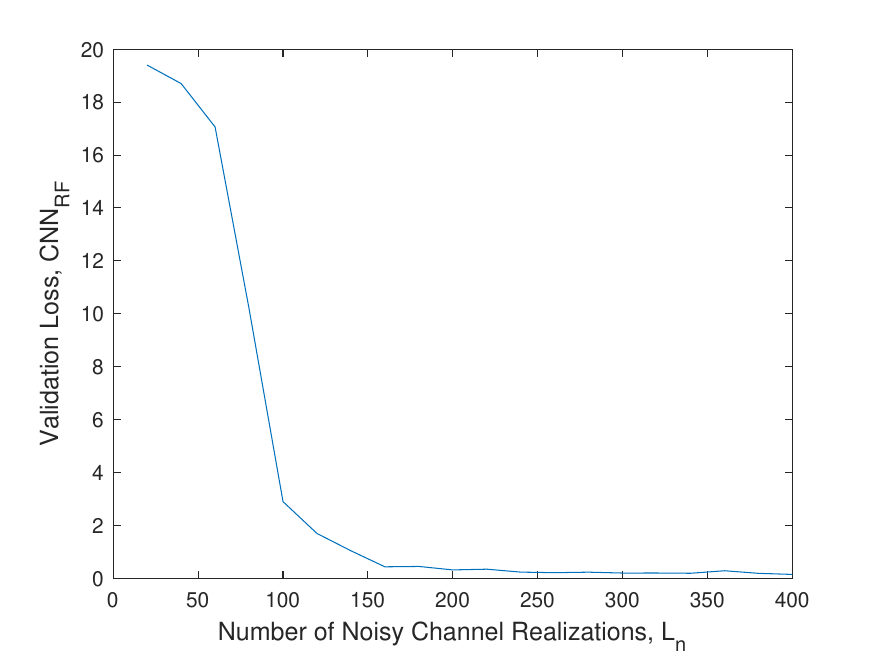}
			 }
		\subfloat[][]{\includegraphics[width=.45\textwidth]{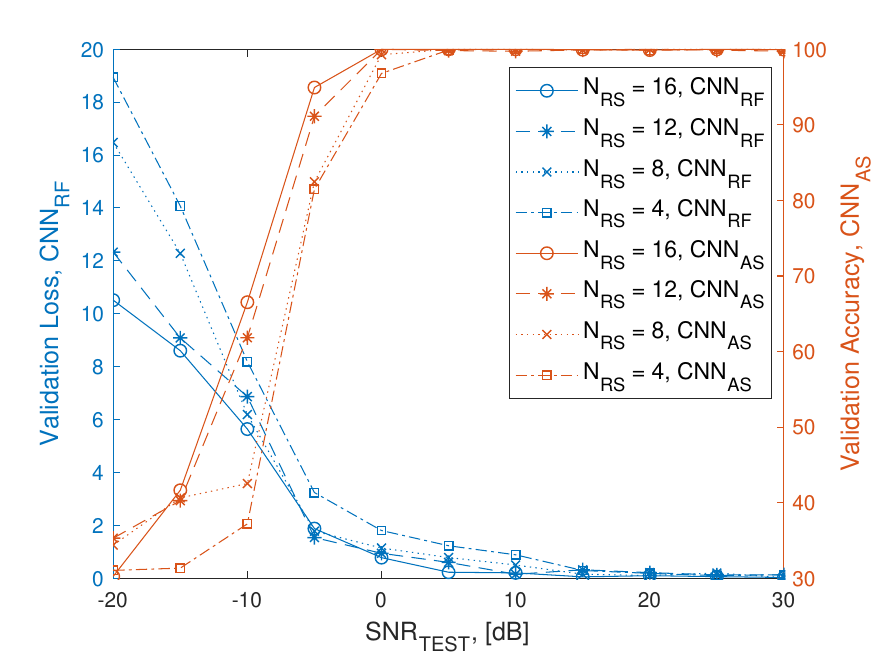}
			 } \\
		\subfloat[][]{\includegraphics[width=.45\textwidth]{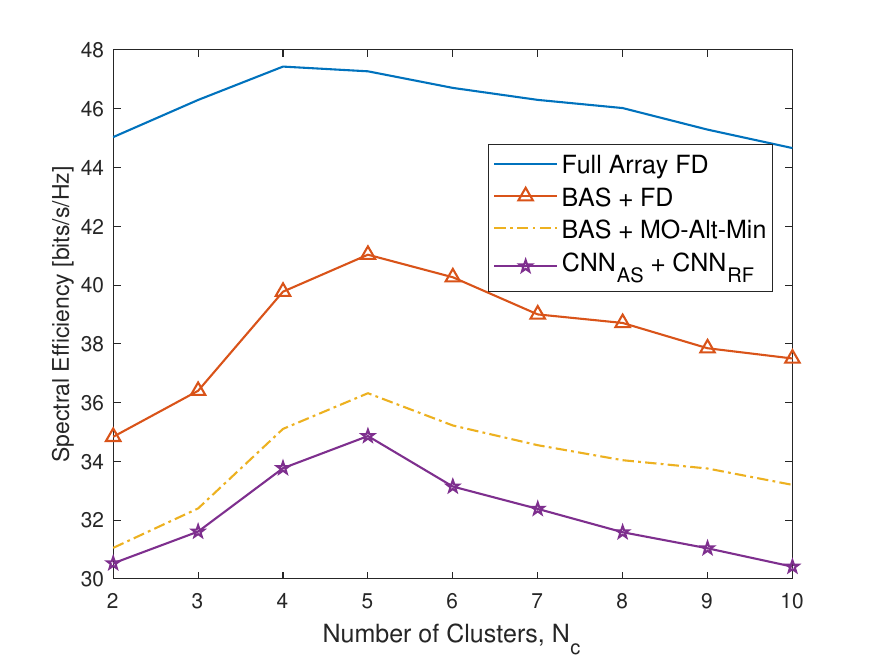}
		}
		\subfloat[][]{\includegraphics[width=.45\textwidth]{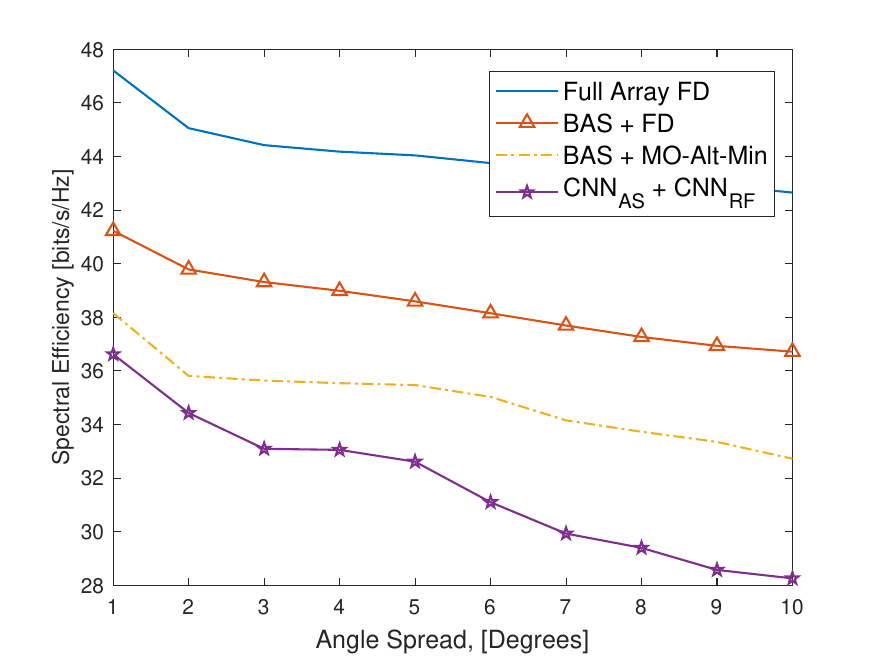}
		 } 
		\caption{Performance of CNN. (a) Validation loss versus number of channel realizations $L_\mathrm{n}$. (b) Validation loss for CNN$_\mathrm{RF}$ and validation accuracy for CNN$_\mathrm{AS}$. (c) Spectral efficiency versus number of clusters in when the angle spread is $\sigma_{\Theta}^2 = 5^\circ$. (d) Spectral efficiency versus angle spread  when the number of paths $N_\mathrm{c} = 4$. We set the number of antennas as $N_T = N_R=64$  with $N_{RS}=16$ selected antennas \cite{elbirQuantizedCNN2019}.}
		\label{figPerformanceOfCNN}
	\end{figure*}

	Figure~\ref{figPerformanceOfCNN} summarizes the assessment of the performance of unquantized CNN. Here, we set $N=100$, $N_T = N_R = 64$ and  $N_S = 4$. Figure~\ref{figPerformanceOfCNN}a shows the validation loss of CNN$_\mathrm{RF}$ against the number of noisy channel realizations $L$. We observe that the loss is satisfactory for $L\geq 150$; in the simulations, we keep $L=200$. Note that this is a common result for different number of channel realizations $N$. In fact, we use $N=100$  to achieve reasonable network accuracy. To investigate the performance of deep networks against different noise levels in the training data, we demonstrate the validation loss of CNN$_\mathrm{RF}$ and the classification accuracy of CNN$_\mathrm{AS}$ in Fig.~\ref{figPerformanceOfCNN}b for different $N_{RS}$ values. It is clear here that both networks attain satisfactory network accuracy for SNR$_\mathrm{TEST} \geq 0$dB. At low SNR regimes, CNN has poor classification performance due to the deviations between the input and the channel matrices used in the training data. In order to make CNN more robust to noisy inputs, we draw the training data for multiple SNR$_\text{TRAIN}$ levels. Nevertheless, noise in the training data expectedly limits the performance since the network cannot distinguish the input data if it is corrupted too much. This issue is also reported in \cite{elbir2019cognitive,elbirDL_COMML} for the multiple SNR$_\text{TRAIN}$ case.
	
	The channel statistics are important parameters that change in very short time in mm-Wave channels. Hence, in the training stage, we feed the network with channel realizations of different $N_\mathrm{c}$ values. To further investigate the performance with respect to different channel statistics, we compare the algorithms in Fig.~\ref{figPerformanceOfCNN}c and d for different $N_\mathrm{c}$ and $\sigma_{\Theta}^2$, respectively. The  CNN framework provides robust performance against different channel statistics. The effectiveness of the DL techniques can be attributed to training the network with several channel statistics and adding synthetic noise for multiple SNR$_\mathrm{TRAIN}$ values.
	
	As a result, the DL network provides robust performance against the changes in channel statistics without a need to be re-trained. However, when there is a change in the \textit{full array} system parameters such as $N_T$, $N_R$, $N_{RS}$ and $N_S$, the network does need to be re-trained because these parameters directly dictates the dimensions of the network input and output layers.



	\begin{figure*}[t]
		\centering
		{\includegraphics[width=1.0\textwidth]{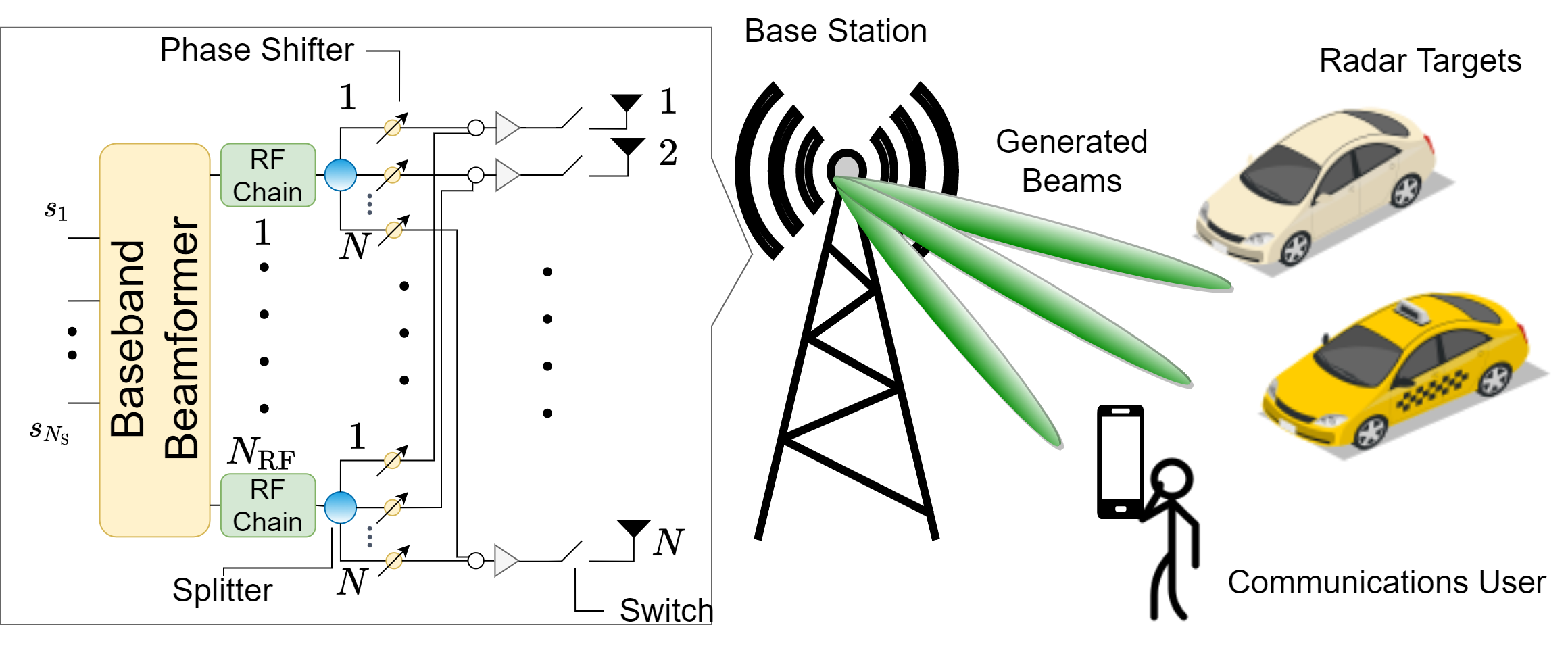}}
		\caption{ISAC hybrid beamforming scenario with antenna selection. }
		\label{fig_ISAC_BS}
	\end{figure*}

\section{Deep Sparse Arrays for ISAC}
\label{sec:isac}
 To jointly access and manage the limited electromagnetic spectrum, ISAC has been envisaged as a technology that will use the same transmit/receive hardware for both radar and communications \cite{mishra2019toward,elbir2021terahertz}. In ISAC (Fig.~\ref{fig_ISAC_BS}), the aim is to select the subarray that provides both the best radar sensing and communications performance. In particular, we design the hybrid beamformers corresponding to the selected subarray while maximizing the beamforming performance toward both radar targets and communications users. Due to the combinatorial nature of the antenna selection problem, the computational complexity of the joint (antenna selection and beamforming) problem is high. To combat this challenging problem, a low-complexity approach is devised, wherein the consecutive antennas in the whole array are partitioned into small groups so that the number of possible subarray configurations can be reduced. Then, we introduce an optimization problem to jointly select the \textit{best} subarray and design the hybrid analog and digital beamformers for the ISAC scenario. Finally, a DL-based approach is proposed to further lower the computational burden of the design problem. In particular, a CNN is devised such that the input of the CNN is the combination of both communications (channel matrix) and sensing (radar-only beamformer) related data. The output of the learning model is the combination of the digital beamformer and the sparse analog beamformer. 
 
\begin{figure*}[t]
		\centering
		{\includegraphics[width=.5\textwidth]{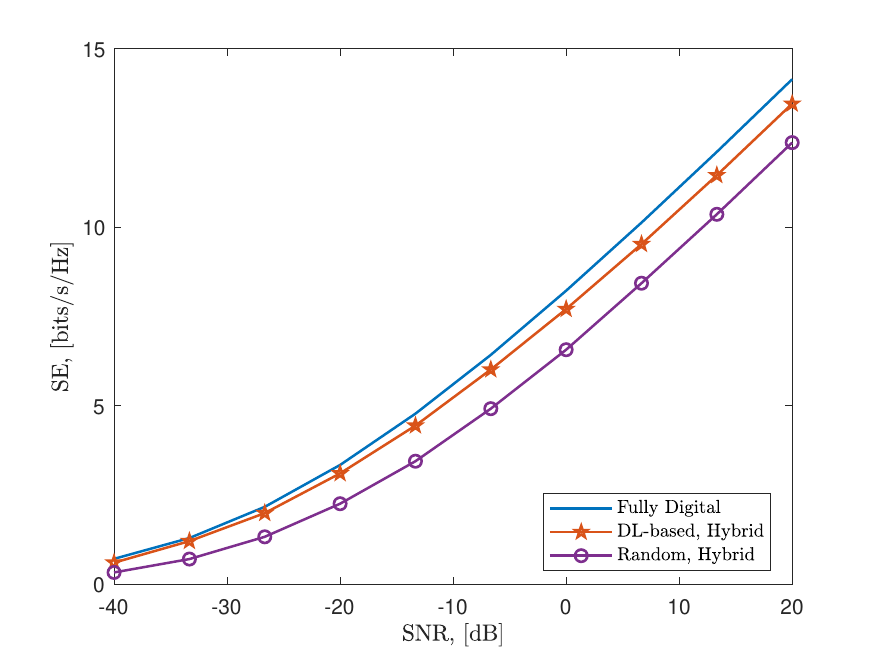}}
		\caption{ISAC hybrid beamforming performance with antenna selection. }
		\label{fig_ISAC_SNR}
	\end{figure*}

 Fig.~\ref{fig_ISAC_SNR} shows the spectral efficiency (SE) of the joint antenna selection and beamformer design in the ISAC scenario. DL-based CNN achieves a performance very close to the fully digital ISAC beamformer~\cite{elbir2021terahertz}, which is computed in accordance with the selected sparse array by the CNN. In comparison, the random selection has a poorer SE. 


\section{Summary}
\label{sec:summ}
 In this chapter, we reviewed DL-based sparse array design principles, procedures, and applications with a focus on direction finding. In general, this problem may be viewed as either a classification or regression in the DL framework. Our DL-based design offers the advantages of feature engineering and prediction-stage computational efficiency. This is particularly useful for dynamic sparse array design scenarios such as in cognitive radar, where a new sparse array must be selected to adapt to the changes in the target environment. Here, we employed a CNN to find the antenna subarray that yields the minimum CRB of the DoA estimation error. Further, antenna selection knowledge already learned at a pre-trained model may be used for a different antenna array geometry using TL methods. This has applications in metacognitive radars \cite{mishra2020toward}. For 2-D or planar sparse arrays, we illustrated the use of SA-assisted DL algorithms. There are a large number of deep sparse array applications in wireless communications, wherein antenna selection is generally employed to reduce the complexity of common functions such as beamforming, channel estimation, and ISAC. 

 There are many interesting open challenges that need to be tackled for practical and futuristic implementations of DL-based sparse array techniques. Training a learning model still requires solving complex optimization problems for each training data sample offline prior to training. For instance, the learning models are pivoted by a reasonable trade-off between an efficient implementation and satisfactory beampattern (radar sensing) and spectral efficiency (communications) performance. Practical implementations may employ compressed or quantized model parameters to reduce model complexity. Further, data/model parallelization is useful for accelerating training times.





\bibliographystyle{vancouver}
\bibliography{main}


\end{document}